\documentclass[fleqn,usenatbib]{mnras}

\usepackage{newtxtext,newtxmath}
\usepackage[english,german,spanish,french]{babel}
\usepackage[T1]{fontenc}

\DeclareRobustCommand{\VAN}[3]{#2}
\let\VANthebibliography\thebibliography
\def\thebibliography{\DeclareRobustCommand{\VAN}[3]{##3}\VANthebibliography}

\usepackage{graphicx}	
\usepackage{newtxtext,newtxmath}
\usepackage{pdflscape}
\usepackage{multirow}
\usepackage{xcolor}
\usepackage{adjustbox,lipsum}
\usepackage{academicons}

\DeclareRobustCommand{\ion}[2]{
\relax\ifmmode
\ifx\testbx\f@series
{\mathbf{#1\,\mathsc{#2}}}\else
{\mathrm{#1\,\mathsc{#2}}}\fi
\else\textup{#1\,{\mdseries\textsc{#2}}}
\fi}

\title[DESI PV target selection]{Target Selection for the DESI Peculiar Velocity Survey}

\author[C. Saulder et al.]{
Christoph Saulder$^{1}$\thanks{E-mail: csaulder@kasi.re.kr}, 
Cullan Howlett$^{2}$, 
Kelly A. Douglass$^{3}$, 
Khaled Said$^{2}$, 
Segev BenZvi$^{3}$, 
\newauthor
Steven Ahlen$^{4}$, 
Greg Aldering$^{5}$, 
Stephen Bailey$^{5}$, 
David Brooks$^{6}$, 
Tamara M.~Davis$^{2}$, 
Axel de la Macorra$^{7}$, 
\newauthor
Arjun Dey$^{8}$
Andreu Font-Ribera$^{9}$, 
Jaime E. Forero-Romero$^{10}$,
Satya Gontcho A Gontcho$^{5}$, 
\newauthor
Klaus Honscheid$^{11,12}$, 
Alex G. Kim$^{5}$, 
Theodore Kisner$^{5}$, 
Anthony Kremin$^{5}$, 
Martin Landriau$^{5}$, 
\newauthor
Michael E. Levi$^{5}$, 
John Lucey$^{13}$, 
Aaron M. Meisner$^{8}$, 
Ramon Miquel$^{9,14}$, 
John Moustakas$^{15}$, 
\newauthor
Adam~D.~Myers$^{16}$, 
Nathalie Palanque-Delabrouille$^{5,17}$, 
Will Percival$^{18,19,20}$,
Claire Poppett$^{5,21,22}$,
\newauthor
Francisco Prada$^{23}$, 
Fei Qin$^{1}$, 
Michael Schubnell$^{24}$,
Gregory Tarl\'{e}$^{24}$, 
Mariana Vargas Maga\~na$^{7}$, 
\newauthor
Benjamin Alan Weaver$^{8}$, 
Rongpu Zhou$^{5}$,
Zhimin Zhou$^{25}$, 
Hu Zou$^{25}$\\
$^{1}$Korea Astronomy and Space Science Institute, 776 Daedeok-daero, Yuseong-gu, 34055 Daejeon, Republic of Korea.\\
$^{2}$School of Mathematics and Physics, University of Queensland, Brisbane, QLD 4072, Australia. \\
$^{3}$Department of Physics \& Astronomy, University of Rochester, 500 Wilson Blvd., Rochester, NY 14627, USA. \\
$^{4}$Physics Dept., Boston University, 590 Commonwealth Avenue, Boston, MA 02215, USA\\
$^{5}$Lawrence Berkeley National Laboratory, 1 Cyclotron Road, Berkeley, CA 94720, USA\\
$^{6}$Department of Physics \& Astronomy, University College London, Gower Street, London, WC1E 6BT, UK\\
$^{7}$Instituto de F\'{\i}sica, Universidad Nacional Aut\'{o}noma de M\'{e}xico,  Cd. de M\'{e}xico  C.P. 04510,  M\'{e}xico\\
$^{8}$NSF's NOIRLab, 950 N. Cherry Ave., Tucson, AZ 85719, USA\\
$^{9}$Institut de F\'{i}sica d’Altes Energies (IFAE), The Barcelona Institute of Science and Technology, Campus UAB, 08193 Bellaterra Barcelona, Spain\\
$^{10}$Departamento de F\'isica, Universidad de los Andes, Cra. 1 No. 18A-10, Edificio Ip, CP 111711, Bogot\'a, Colombia\\
$^{11}$Center for Cosmology and AstroParticle Physics, The Ohio State University, 191 West Woodruff Avenue, Columbus, OH 43210, USA\\
$^{12}$Department of Physics, The Ohio State University, 191 West Woodruff Avenue, Columbus, OH 43210, USA\\
$^{13}$Centre for Extragalactic Astronomy, Department of Physics, Durham University, South Road, Durham DH1 3LE, UK\\
$^{14}$Instituci\'{o} Catalana de Recerca i Estudis Avan\c{c}ats, Passeig de Llu\'{\i}s Companys, 23, 08010 Barcelona, Spain\\
$^{15}$Department of Physics \& Astronomy, Siena College, 515 Loudon Rd., Loudonville, NY, 12211, USA. \\
$^{16}$Department of Physics \& Astronomy, University  of Wyoming, 1000 E. University, Dept.~3905, Laramie, WY 82071, USA\\
$^{17}$IRFU, CEA, Universit\'{e} Paris-Saclay, F-91191 Gif-sur-Yvette, France\\$^{16}$Department of Physics \& Astronomy, Siena College, 515 Loudon Rd., Loudonville, NY, 12211, USA. \\
$^{18}$Department of Physics and Astronomy, University of Waterloo, 200 University Ave W, Waterloo, ON N2L 3G1, Canada\\
$^{19}$Perimeter Institute for Theoretical Physics, 31 Caroline St. North, Waterloo, ON N2L 2Y5, Canada\\
$^{20}$Waterloo Centre for Astrophysics, University of Waterloo, 200 University Ave W, Waterloo, ON N2L 3G1, Canada\\
$^{21}$Space Sciences Laboratory, University of California, Berkeley, 7 Gauss Way, Berkeley, CA  94720, USA\\
$^{22}$University of California, Berkeley, 110 Sproul Hall \#5800 Berkeley, CA 94720, USA\\
$^{23}$Instituto de Astrof\'{i}sica de Andaluc\'{i}a (CSIC), Glorieta de la Astronom\'{i}a, s/n, E-18008 Granada, Spain\\
$^{24}$Department of Physics, University of Michigan, Ann Arbor, MI 48109, USA\\
$^{25}$National Astronomical Observatories, Chinese Academy of Sciences, A20 Datun Rd., Chaoyang District, Beijing, 100012, P.R. China
}

\date{Accepted XXX. Received YYY; in original form ZZZ}

\pubyear{2022}

\begin{document}
\label{firstpage}

\maketitle

\begin{abstract}
We describe the target selection and characteristics of the DESI Peculiar Velocity Survey, the largest survey of peculiar velocities (PVs) using both the fundamental plane (FP) and the Tully-Fisher (TF) relationship planned to date. We detail how we identify suitable early-type galaxies (ETGs) for the FP and suitable late-type galaxies (LTGs) for the TF relation using the photometric data provided by the DESI Legacy Imaging Survey DR9. Subsequently, we provide targets for 373 533 ETGs and 118 637 LTGs within the DESI 5-year footprint. We validate these photometric selections using existing morphological classifications. Furthermore, we demonstrate using survey validation data that DESI is able to measure the spectroscopic properties to sufficient precision to obtain PVs for our targets. Based on realistic DESI fiber assignment simulations and spectroscopic success rates, we predict the final DESI PV Survey will obtain $\sim$133 000 FP-based and $\sim$53 000 TF-based PV measurements over an area of 14 000 $\mathrm{deg^{2}}$. 
We forecast the ability of using these data to measure the clustering of galaxy positions and PVs from the combined DESI PV and Bright Galaxy Surveys (BGS), which allows for cancellation of cosmic variance at low redshifts. With these forecasts, we anticipate a $4\%$ statistical measurement on the growth rate of structure at $z<0.15$. This is over two times better than achievable with redshifts from the BGS alone. The combined DESI PV and BGS will enable the most precise tests to date of the time and scale dependence of large-scale structure growth at $z<0.15$.
\end{abstract}




\section{Introduction}
\label{sec:intro}


The Dark Energy Spectroscopic Instrument (DESI) is currently carrying out a 5-year survey of $\sim$40 million redshifts over $\sim$14 000 deg$^{2}$ with the aim of measuring the expansion and growth history of our Universe to unprecedented precision \citep{DESI_white_new}. This will be achieved primarily by measuring the clustering of multiple galaxy types between redshifts $0.0 < z < 3.5$ and utilising the well-understood signatures of Baryon Acoustic Oscillations (the remnants of sound waves frozen into the galaxy distribution close to the epoch of recombination, \citealt{Eisenstein1998}) and Redshift Space Distortions (RSD; the apparent anisotropy in the clustering of galaxies due to the use of observed redshifts to infer galaxy positions, \citealt{Kaiser:1987}).

DESI uses a series of ten fibre-fed spectrographs that are able to take up to 5000 spectra over a wavelength range from 360 nm to 980 nm simultaneously \citep{DESI_instrument}. DESI is installed at the 4-m Mayall telescope in Kitt Peak, Arizona. The main survey will observe samples of Bright (BGS, \citealt{DESI_target_BGS}), Luminous Red (LRG, \citealt{DESI_target_LRG}), and Emission Line Galaxies (ELG, \citealt{DESI_target_ELG}), Milky Way stars (MWS, \citealt{DESI_target_MWS}), and Quasi-Stellar Objects (QSO, \citealt{DESI_target_QSO} and their Lyman-$\alpha$ forest). Observations are divided into dark and bright time in order to maximize the number of spectra obtained within the five year survey considering varying observing conditions, with the fainter target classes (LRG, ELG, and QSO) only observed under the best conditions (dark time) and the brighter target classes (BGS and MWS) otherwise (bright time).

In addition, DESI will also have a number of spare fibres, particularly due to the design of its focal plane \citep{DESI_focalplane} as well as its nature as a multi-pass survey. Following an internal review process, a number of secondary target programs were identified that make use of these spare fibres and add value to, or augment, the main survey goals. The DESI peculiar velocity (PV) survey is one such program, and the subject of this work.

The observed, CMB-frame redshift\footnote{CMB-frame redshift refers to the observed redshift after correction for the motion of the telescope with respect the Cosmic Microwave Background. This correction is usually performed with separate terms accounting for the rotation of the Earth, the orbital motion of the Earth, and the motion of the Solar System with respect to the CMB as measured by \cite{COBE} or \cite{Planck2018}} of a galaxy contains contributions from the expansion of the Universe $z_{\mathrm{cosmo}}$ and the motion of the observed galaxy due to the gravitational forces from nearby structures $z_{\mathrm{p}} = v_{\mathrm{p}}/c$. $v_{\mathrm{p}}$ is the PV, which can be measured by combining the observed redshift with a distance indicator that enables us to infer $z_{\mathrm{cosmo}}$ independently (e.g., \citealt{Davis:2014} and \citealt{Watkins:2015}). The PVs of galaxies are typically in order of a few hundred km s$^{-1}$. 

Redshift Space Distortions (RSD) statistically probe the PVs of galaxies as a departure from the otherwise homogeneous and isotropic expansion of the Universe and provide an opportunity to test the predictions of General Relativity (GR) by measuring the rate at which large scale inhomogeneities grow. Typically these are parameterised by the ``growth rate of structure'' $f(z)$, which is usually normalised  and expressed as
\begin{equation}
f(z)\sigma_8(z)=\Omega^{\gamma}_{m}(z)\sigma_{8}(z), 
\label{eq:fsigma8}
\end{equation}
where $\gamma$ is the growth index, $\Omega_{m}$ is the relative matter density of the universe, and $\sigma_{8}$ stands for the amplitude of the linear power spectrum on the scale of $8 h^{-1}$ Mpc. One of the main goals of DESI is to place tight constraints on potential large-scale deviations from GR, modified gravity theories and Dark Energy using measurements of RSD over the last 10 billion years.

Surveys that directly measure PVs have demonstrated dramatic improvement on growth of structure constraints relative to RSD alone at $z \lesssim 0.2$, both theoretically (e.g., \citealt{Burkey2004, Irsic:2011, Koda2014, Howlett2017a, Howlett2017b, Whitford2021}), and practically (e.g., \citealt{Carrick2015, Qin2019, Adams2020, Said:2020, Lai2022}). This is because joint measurements of the density and velocity fields can cancel out the effects of sample variance in a similar way to a multi-tracer RSD analysis \citep{McDonald2009b}. PVs also directly probe the underlying matter field, independent of galaxy bias, and are sensitive to larger scale modes than the density field. These effects make PVs particularly useful at low (z < $\sim0.1$) redshifts, where the volume of the Universe that can be surveyed is small. At these redshifts, PVs can be measured most accurately, and various theories of modified gravity present the largest differences in phenomenology. Coupled with redshift surveys, PVs can be used to map the cosmography of the local universe \citep{Courtois:2013, Tully2014, Graziani2019}, and to correct the redshifts of transients before using them to build the Hubble diagram \citep{Howlett2020, Carr2021, SH0ES, Pantheon_plus}.

In order to measure PVs, one needs a distance indicator. There are many such indicators, including Cepheid variable stars \citep{Leavitt1912}, the tip of the red giant branch \citep{Lee1993}, Type Ia supernovae \citep{Phillips1993}, surface brightness fluctuations \citep{Tonry1988}, the Tully-Fisher relation (TF; \citealt{Tully_Fisher}), the Fundamental Plane (FP; \citealt{Djorgovski:1987, Dressler:1987}), and mergers of compact objects detected via gravitational waves \citep{Holz2005,Palmese2021}. The two currently most suited for measuring the distances to large numbers of galaxies (on the order of several 100 000 as we will show in this work) are the FP and the TF relation. These are what will be used in the DESI PV survey. However, unlike previous PV surveys which usually use one or the other of these two probes (e.g., \citealt{Hong:2019,Springob:2007,Springob:2014,Howlett2022}), the DESI PV survey will combine these two distance indicators in an unprecedented way using consistent target selection methods for both indicators and the same instrument across the entire DESI footprint. 
A downside of these methods is that with distance, FP and TF uncertainties grow in proportion to redshift, which limits their use to low redshifts. However, the sizeable ($\sim25$\%) uncertainty for FP and TF distance indicators is ameliorated by the large numbers that DESI will be able to observed.


The aim of this paper is to present and validate the DESI PV survey. We start in Section~\ref{sec:methods} with a description of the targets that will be observed and how they were selected. In Sections~\ref{sec:results_FP} and Sections~\ref{sec:results_TF} we justify our selection using data for a few objects obtained during DESI's Survey Validation (SV) program and present a proof-of-concept that DESI will be able to accurately measure PVs using the FP and the TF relation. We then provide forecasts for what the DESI PV survey will produce after 5 years of operation in Section~\ref{sec:survey}, including sky area and predicted numbers of successful PV measurements. We then propagate these through into cosmological parameters in Section~\ref{sec:forecasts}. We conclude in Section~\ref{sec:conclusions}. For all forecasts in this work, we adopt a \textit{Planck}-based cosmology \citep{Planck_cosmo} with $\Omega_{m,0} = 0.3121$, $\Omega_{b,0} = 0.0488$, $H_{0} = 100h = 67.51\,\mathrm{km\,s^{-1}\,Mpc^{-1}}$, $n_{s} = 0.9653$, and $\sigma_{8}(z=0)=0.815$.

\section{Target Selection}
\label{sec:methods}

The DESI PV survey consists of two main distance indicators/PV tracers: early-type galaxies calibrated using the FP, and spiral galaxies calibrated using the TF relation.

The FP, which was properly defined and discussed in \citet{Dressler:1987} and \citet{Djorgovski:1987}, after being first mentioned in \citet{Terlevich:1981}, is an empirical relation between three directly observable parameters of elliptical galaxies: the angular effective radius, $\theta_{e}$, the effective surface brightness, $I_{e}$, and the central velocity dispersion $\sigma$. For a given set of cosmological parameters and the angular diameter distance derived using them, the angular effective radius can be converted to a physical radius, $R_{e}$. The FP is then expressed as 
\begin{equation}
\log(R_{e}) = a\log(\sigma) + b\log(I_{e}) + c.
\label{eq:FP}
\end{equation}

Given the assumption that a sample of galaxies all lie along the same plane given by Eq.~\ref{eq:FP}, the FP can be used as a distance indicator by comparing the measured and predicted angular sizes for each galaxy given their observed redshifts and values of $I_{e}$ and $\sigma$. The inferred PV is hence proportional to the offset of each galaxy from the plane.
However, the uncertainties are typically large due to our inability to disentangle the contributions to the total offset from the PV and intrinsic, astrophysical, scatter in the FP relationship itself. Calibrations and applications of the FP have been previously carried out for other large-scale spectroscopic surveys such as SDSS \citep{SDSS_DR15} and the 6dFGS \citep{6dFGS_final} with notable work by \citet{Bernardi:2003c,Hyde:2009,Magoulas:2012,Saulder:2013,Campbell:2014,Qin:2018,Said:2020,Howlett2022}.

The second distance indicator used here, the TF relation, is the empirical relationship between rotational velocity $V_{\mathrm{max}}$ and the absolute magnitude $M$ of spiral galaxies. This correlation was first used by \cite{Opik1922} during the Great Debate to prove that Andromeda is an extragalactic object. More than 50 years later, \cite{Balkowski1974} described this correlation using a sample of spiral and irregular dwarf galaxies without using it as a distance indicator. \citet{Tully_Fisher} were the first to propose the linear relation 
\begin{equation}
  M = b\log(V_{\text{max}}) + c,
\end{equation}
as a distance indicator for inclined spiral galaxies. In this methodology, the measured apparent magnitude $m$ of a set of template galaxies is measured and combined with known distances to evaluate the absolute magnitude. The resulting best-fit TF relation is then used to estimate the absolute magnitude for a second sample which measured apparent magnitudes, from which we can then infer the distance. The PV is hence proportional to the offset from the TF relation, but again has sizeable errors due to the unknown amplitude of the intrinsic astrophysical scatter ($\sim$20\%) in the relationship. 

The arrival of CCD detectors made this relation more easily accessible to cosmological community \citep{Pierce1988,Courteau1993} and was used to carry out most of the PV surveys in the local universe such as the Spiral Field and Cluster I-band TF  (SFI++; \citealt{Springob2007}), CosmicFlows (\citealt{Tully2009,Kourkchi:2020}), and the 2MASS Tully-Fisher Survey (2MTF: \citealt{Masters:2008,Hong:2019}). 

The DESI PV survey will produce samples of FP and TF galaxies using imaging data (to compute $\theta_{e}$, $I_{e}$ and $m$) combined with new spectroscopic measurements (to obtain the redshift, $V_{\mathrm{max}}$ by comparing redshifts measurements at different positions along the semi-major axis of TF galaxies, and the velocity dispersion $\sigma$ by fitting the line widths of FP galaxies). However, before a spectroscopic survey can be launched one needs to know where to point the fibres. To this end, a large imaging survey was carried out as a prelude to DESI, the DESI Legacy Imaging Survey \citep{DESI_Imaging}. The ninth dataset (DR9) of this survey was used for target selection for the DESI PV survey (and other DESI main surveys), and will be used to produce the aforementioned photometric properties for the FP and TF samples.

The DESI Legacy Imaging Survey combines data from three different telescopes covering about half the sky. The observations for the northern part of the imaging catalogue (declination greater than $32^\circ$) were carried out in the $g$ and $r$ bands as the Beijing-Arizona Sky Survey (BASS) using the Bok telescope, supplemented by $z$-band images from the Mayall z-band Legacy Survey (MzLS), which was carried out on the 4m Mayall telescope. For the sky further south, the Dark Energy Camera on the Blanco 4m telescope was used for the Dark Energy Camera Legacy Survey (DeCALS) and the well-known Dark Energy Survey \citep{DES_3Y_photo}. Additionally, the data is supplemented with matched infrared photometry from WISE \citep{WISE_mision} using the five-year unWISE Coadds \citep{unWISE:5Y}. The DESI Legacy Imaging Survey has already been used to obtain derived data products such as the Siena Galaxy Atlas\footnote{The SGA catalogue can be found on: {\url{https://www.legacysurvey.org/sga/sga2020/}}} (SGA; \citet{MoustakasSGA}) for galaxies with large angular radii, and photometric redshifts \citep{DESI_photoz}. We take advantage of both of these in our own target selection, detailed in the rest of this Section.

\subsection{Basic sample selection}
\label{sec:sample}
Our selection starts with the sweep catalogues\footnote{\url{https://portal.nersc.gov/cfs/cosmo/data/legacysurvey/dr9/north/sweep/} and \url{https://portal.nersc.gov/cfs/cosmo/data/legacysurvey/dr9/south/sweep/}} of the DESI Legacy Imaging Surveys DR9 \citep{DESI_Imaging}, plus the photometric redshift catalogues of \cite{DESI_photoz} and the SGA \citep{MoustakasSGA}.

Our initial cuts match the preliminary definition of the Bright Galaxy Survey (BGS; \citealt{BGS_pre_target}). We required that each object have a successful source detection in all bands ($g$, $r$, and $z$ band). Magnitude limits in both model magnitude and fibre magnitude ensured that only galaxies with a good chance of successful spectroscopic measurements are included. Masks were employed to remove bright stars, globular clusters, and large galaxies (objects in the SGA) from the preliminary BGS sample. Star-galaxy separation using GAIA \citep{GAIA_DR2} data and colour cuts was used to remove obvious stars from the data set. Additional cuts in galaxy colours and fraction of flux from overlapping sources were imposed to insure quality. Pseudo-code for the preliminary BGS target selection is given in Appendix \ref{app:BGS}.

As a second step, we supplemented this preliminary BGS sample with all non-point sources that are brighter than 18 mag in the $r$ band as well as the entire SGA sample.\footnote{In the preliminary definition of the BGS sample, the large galaxies of the SGA were excluded. However, in the final rendition, which was defined after we had set our selection criteria, the SGA was reintroduced as part of the BGS. This does not affect the overall targeting for DESI as duplicate targets are merged and assigned flags that allow us to identify the same target as of interest to both the final BGS definition and PV survey. In practice, this just means that more of our targets will already be observed as part of the DESI main program rather than requiring spare fibres.} However, we then focus on a subsample of the BGS by removing sources that we deemed too faint to obtain sufficient quality data to be used for the FP or TF relation or too distant to obtain accurate PVs (see Section~\ref{sec:results} for further details). 

To this end, we required our potential targets to have a photometric redshift (based on \citealt{DESI_photoz}) of less than 0.15 with the lower $95\% $ confidence interval of the posterior being below 0.1. Because of the $\sim20\%$ intrinsic scatter in the FP and TF relation, typical PV errors for these galaxies scale as $0.2 c z_{\mathrm{obs}}$. At $z_{\mathrm{obs}}=0.1$, this corresponds to an uncertainty of $\sim6000$ km s$^{-1}$ on a measurement we expect (from theoretical arguments) to be on the order of a few hundred km s$^{-1}$. Hence, our photometric redshift selection ensures that galaxies that would have overwhelmingly large PV errors do not enter the sample, while also being generous in accounting for errors in the photo-$z$'s themselves and our future ability to reduce the intrinsic scatter in the distance indicators. Overall, this allows the targeting of only objects that will likely yield usable results for our science goals. 

We applied our selections to both the North and South footprints of the DESI Legacy Imaging Survey. For the overlapping areas we considered all objects with either a Galactic latitude less than or equal to zero, or a declination less than or equal to 32.375 degree as part of the Southern footprint, while all other objects are considered to be part of the Northern footprint. We also consider all objects with a declination greater than $-30$ degree as potential PV targets. This extends further South beyond the planned footprint of the main DESI surveys, which stop around declination $-18$ degree, but we considered it beneficial to have supplementary targets already defined outside the main footprint over the course of DESI's 5-year timeline. The full data of this basic sample can be found in Tables \ref{tab:base_targets},  \ref{tab:photo_targets}, and \ref{tab:flag_targets}. We did not account for differences in photometric measurements, such as magnitudes and sizes, between the northern part of the imaging catalogue and its southern part. However, some small inconsistencies between the two parts of the imaging survey were found after the selection had already been applied to spectroscopic survey. The differences are sufficiently small to be simply accounted for in the final analysis. These issues will be thoroughly explored and discussed in upcoming papers on the individual samples derived from our selection such as Said et al., in preparation. 

\subsection{Photometric Selection of Early-Type galaxies}
\label{sec:ETG}

\begin{figure*}
\includegraphics[width=\columnwidth]{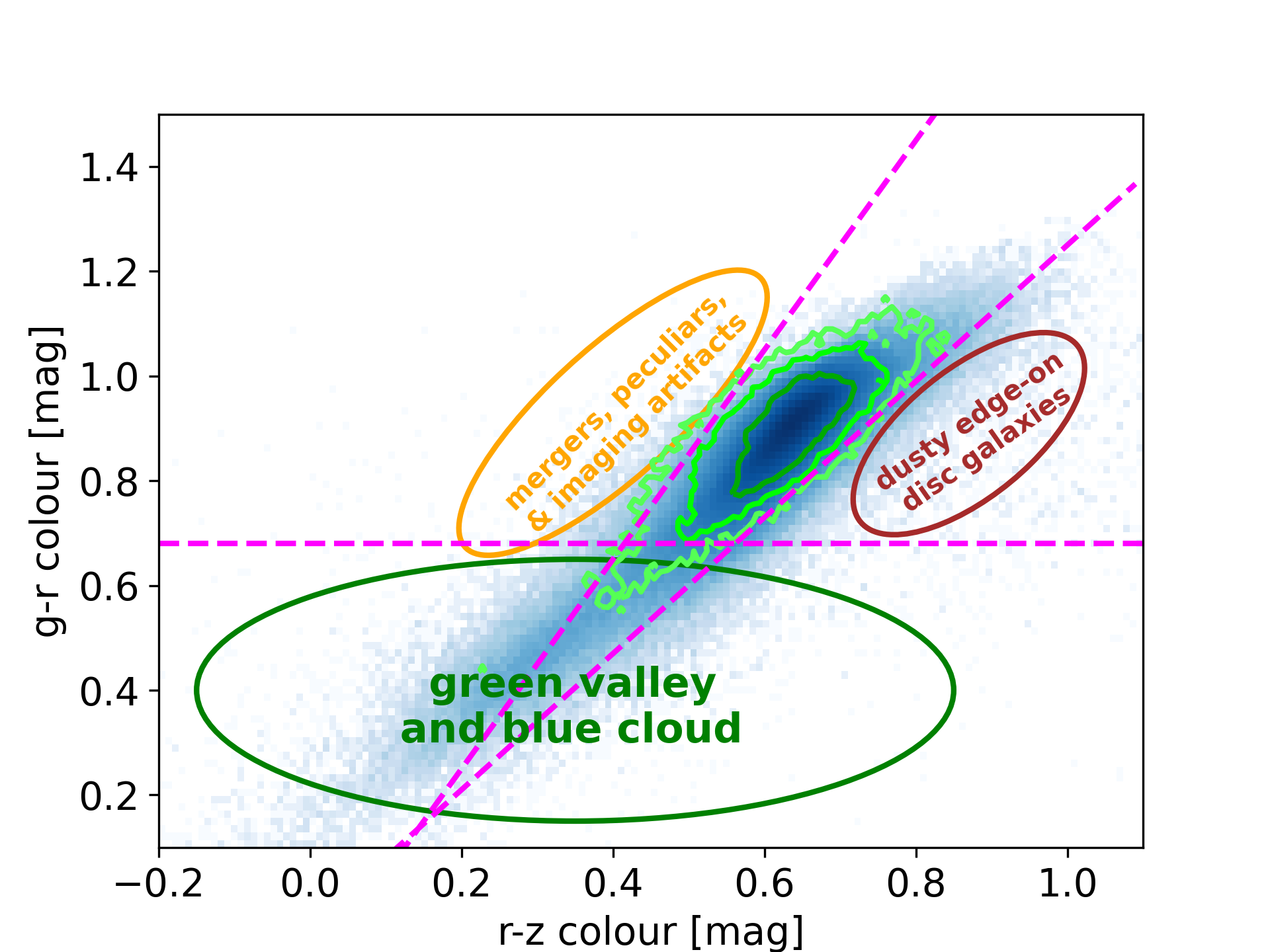}  \includegraphics[width=\columnwidth]{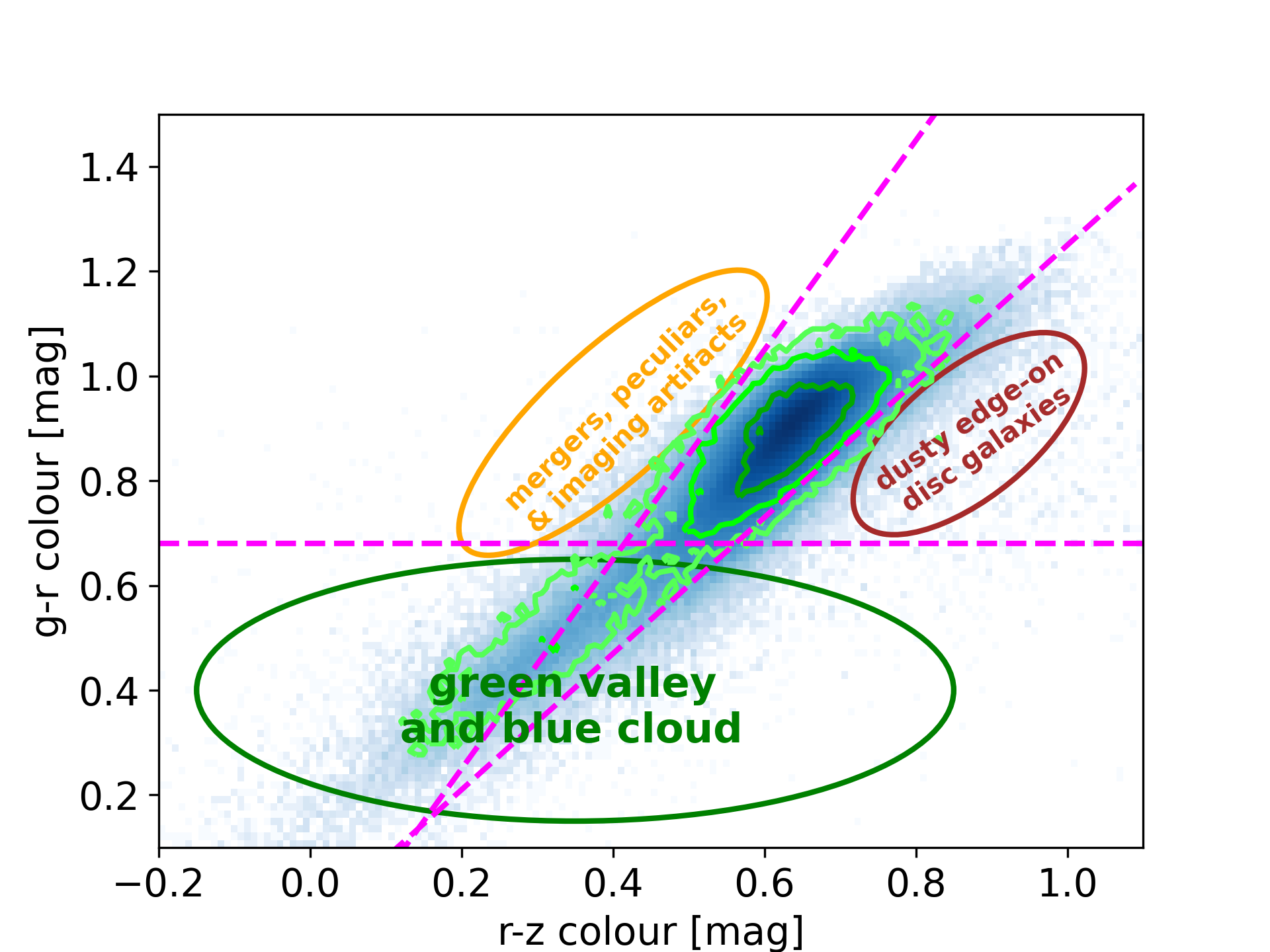}
 \caption{Colour-cuts for the selection of ETGs shown as the region between the two diagonal magenta dashed lines and above the horizontal magenta line, compared to our ETG targets before colour cuts (density map in blue). In the the left panel, the green contours show a truth catalogue based on GalaxyZoo. In the right panel the green contours show a truth catalogue based on the SGA.}
 \label{fig:colourcuts}
\end{figure*}

We build on the basic selection outlined in the previous section to define a series of target sub-classes. The first of these is the FP sample. The full selection process of early-type galaxies (ETG) that will be used to obtain FP distances is two-fold. First we select likely ETGs using a series of simple cuts so that they can be targeted by DESI spare fibres. However, this selection is not perfect. There are interlopers that pass these simple cuts and so need to be removed by more thorough visual inspection. There are also galaxies that photometrically appear to be elliptical, but once spectra are obtained contain non-negligible amounts of H$\alpha$ emission \citep{Bruzual:1998,Blanton:2009,Gomes:2016}. This indicates recent star-forming activity and is likely to be an outlier from the FP (Said et al, in prep.). 

We therefore will need to further refine our selection once spectroscopic measurements have been taken, to remove galaxies that are not suitable for our FP based PV measurements. In this work, we deal with only the photometric selection of ETGs. We validate this selection and make an estimate of the fraction of our objects that would be removed after spectra are obtained, but a more detailed version of the analysis using DESI spectroscopic data will be presented in Said. et. al., (in prep.).

Our goal is to consistently identify nearby ETGs over the entire DESI spectroscopic footprint, while maintaining a low misidentification rate in order to collect a large spectroscopic sample in light of the limited number of spare fibers available on each tile. After extensive tests, on which we elaborate further in Section \ref{sec:results_FP}, using the visual morphological identifications from the SGA \citep{MoustakasSGA} as well as the identifications from GalaxyZoo \citep{GalaxyZoo,GalaxyZoo_data} to calibrate our method, we settled on a set of simple colour cuts and parameters from photometric profile fitting to identify ETG. The morphological classifications and samples based on SGA and GalaxyZoo are discussed in Section \ref{sec:results}. The details of selection criteria for ETGs is provided in this list in pseudo-code:
\begin{itemize}
	\item mag\_r $<$ 18 mag
	\item (mag\_g $-$ mag\_r) $>$ 0.68 mag 
	\item (mag\_g $-$ mag\_r) $>$ (1.3 (mag\_r $-$ mag\_z) $-$ 0.05)
	\item (mag\_g $-$ mag\_r) $<$ (2.0 (mag\_r $-$ mag\_z) $-$ 0.15) 
	\item R\_circ $>$ 0 
	\item (1 $-$ b/a) $<$ 0.7 
	\item (TYPE='DEV') OR (TYPE='SER' AND $n_{s}>2.5$)
	\item $z_{\textrm{photo,median}}$ $<$ 0.15 
	\item $z_{\textrm{photo,low95}}$ $<$ 0.1
\end{itemize}
The variables mag\_g, mag\_r, and mag\_z stand for the extinction corrected magnitudes in each band. R\_circ is the circularised radius and b/a is the semi-minor to semi-major axis ratio. $n_{s}$ is used for the S\'ersic index of the best model fit, which is measured jointly using all optical bands. $z_{\textrm{photo,median}}$ is the median of the photometric redshift and $z_{\textrm{photo,low95}}$ the corresponding lower 95\% boundary confidence of the photometric redshift estimates. 

\begin{figure}
 \includegraphics[width=\columnwidth]{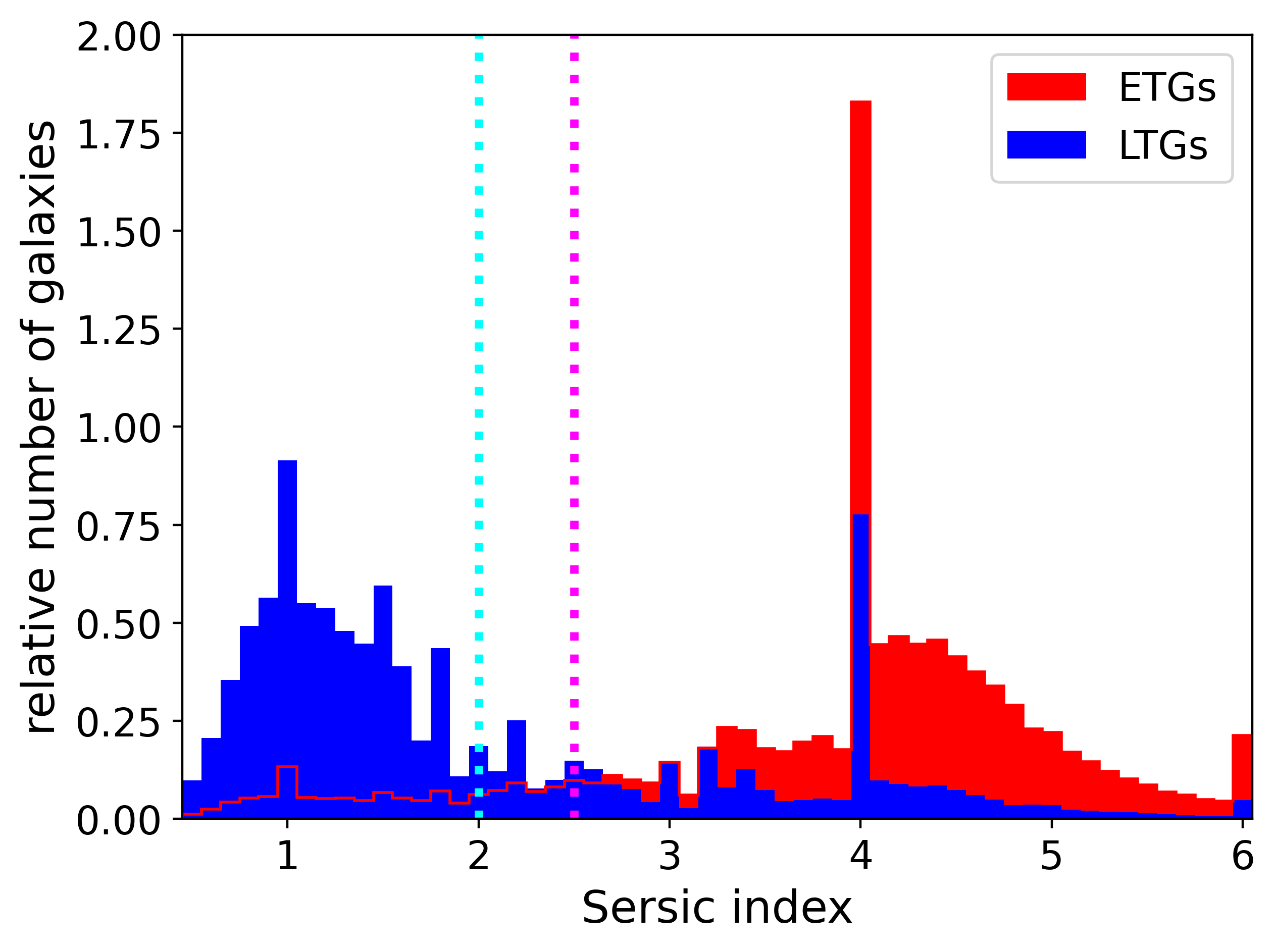}
 \caption{A histogram as a function of S\'ersic index for ETGs and LTGs according to the SGA morphological classifications. The dotted cyan line shows the upper limit of our LTG selection, while the dotted magenta line shows the lower limit of our ETG selection.}
 \label{fig:Sersic_SGA}
\end{figure}

In short, we required that the best fit surface brightness profile of the objects is either a de Vaucouleurs profile or S\'ersic profile with a S\'ersic index greater than 2.5 (see Figure \ref{fig:Sersic_SGA} for the distribution of S\'ersic indices by morphological type according to SGA). Then, a cut on circularised radius larger than zero removed potential unreliable size and axis ratio measurements. We used another cut to remove galaxies with axial ratio $b/a \le 0.3$ as these are typically (close to) edge-on views of lenticular galaxies (S0). The next part of our selection criteria, which required detailed tests and calibrations, was a series of colour cuts, as illustrated in Figure \ref{fig:colourcuts}. A colour cut $g-r > 0.68$ removed galaxies that were below the red sequence, while two other cuts utilising both $g-r$ and $r-z$ colours removed regions of the colour-colour diagram that are plagued by on the one hand dusty galaxies and on the other hand mergers, peculiar objects, and galaxies with imaging artifacts. We further refine our selection by only targeting objects with $r$-band magnitude brighter than 18 mag in order to ensure a reasonable success rate of our velocity dispersion measurements. 

Overall, with the above photometric selection, we identify 427 273 FP galaxies. Each FP galaxy is represented as a single high priority target at the centre of the galaxy. 373 533 of these galaxies are within the tiles covered by the planned 5-year DESI footprint of $\sim$14 000 deg$^{2}$. A complete list of these targets is provided in electronic form as outlined in Table \ref{tab:target_fp}.

\subsection{Photometric Selection of (inclined) Late-Type galaxies}
\label{sec:LTG}
Our second major sub-class of targets is a set of spiral galaxies (late-type galaxies: LTG) suitable for fitting the TF relation. For the photometric selection of these galaxies we again rely on a set of simple criteria, which can be found in pseudo-code in the following list:
\begin{itemize}
\item b/a $<$ cos(25\textdegree) 
\item (TYPE='EXP') OR (TYPE='SER' AND $n_{s}<2$)
\item D$_{26,\textrm{LEDA}} \geq$ 20 arcsec
\item R > 0
\end{itemize}
The variable D$_{25,\textrm{LEDA}}$ is the diameter within which the mean surface brightness exceeds 25 mag/arcsec$^{2}$ according to the LEDA database \citep{LEDA} and R is the uncorrected scale radius of the best profile fit. The thereby defined sample will not cover any object within FP sample as those samples are mutually exclusive.

First of all, in addition to the basic selection criteria in \ref{sec:sample},  we require that the best fit photometric model is either an exponential profile or a S\'ersic profile with a S\'ersic index $n \leq 2$ (see Figure \ref{fig:Sersic_SGA}), isolating disky spiral galaxies. Furthermore, we limit our selection to those galaxies within the SGA sample; all objects therefore have a diameter $D(26) > 20"$; large enough that we are able to place multiple distinct fibers along the major axis of the galaxy. In addition, because galaxies that are close to face-on yield less reliable rotational velocity measurements and therefore are not useful for the TF relation, we require a minimum inclination angle of 25$^\circ$. 

Overall, we identify 129 772 galaxies (118 637 of these galaxies are within the tiles covered by the planned 5-year DESI footprint) that pass our TF selection criteria. As discussed in Section~\ref{sec:results_TF}, we place three targets on each galaxy: one on the centre and two along the semi-major axis, at either $\pm 0.33 R_{26}$ (during SV3) or $\pm 0.4 R_{26}$ (during the main DESI survey) on both sides, making for a total of 389 316 targets.  $R_{26}$ is the semimajor axis radius measured at the $\mu = 26$ mag arcsec$^{-2}$ $r$-band isophote as provided in the SGA. A complete list of these targets is provided in electronic form as outlined in Table \ref{tab:target_tf}. Additionally the parameters used from the SGA for this target selection can be found in Table \ref{tab:sga_targets}.

\subsection{Additional non-primary targets}
\label{sec:exteneded}

In addition to our primary target sub-classes consisting of FP and TF galaxies, we include a set of lower priority targets that provide additional usage for DESI fibers that may otherwise have no science-worthy target (instead being placed as a sky fiber). There are two types: smaller SGA galaxies that do not pass our photometric selections as FP or TF galaxies, and extended objects that may be part of either, or neither, of our primary selections, but are large enough to subtend multiple DESI fiber patrol radii and so may frequently have a DESI fiber land exclusively on them.  

\subsubsection{Additional targets from SGA}
\label{sec:extraSGA}

There are many SGA galaxies that do not make it into the sample of ETGs for the FP or in the sample of late-type galaxy for the TF relation and are not larger than a single DESI fibre patrol radius. Because of the inherent size requirement for inclusion in the SGA, these are typically lenticular galaxies, face-on spirals, or objects with unusual colours. These galaxies, like all SGA objects, are already targets during Bright time, however additional higher quality spectra during dark time are interesting from a galaxy physics point of view. Therefore, although they are not useful for PV science we added them to our target list for the sake of completeness, but at the lowest priority. Each of these SGA galaxies will get the chance for an supplementary dark time observation at it's centre, making for a total of 81 611 additional targets. A complete list of these targets is provided in electronic form as outlined in Table \ref{tab:target_sga}. 

\subsubsection{Extended objects}

Due to the limited patrol radius of each DESI fiber, there is a subset of galaxies that will subtend the entire fiber patrol radius, and thus be the only target available. So that these fibers are not wasted, we target locations in steps of $0.2 R_{26}$ along the semi-major axes of all SGA objects with $D(26) \geq 1.4'$ (the DESI fiber patrol radius), irrespective of morphology.  This comprises 2 911 galaxies. Many of these are also part of our FP or TF samples and so we already target them at their primary locations; but with this step they also are targeted at additional locations, at lower priority. We expect such observations to be useful for systematic tests of our FP and TF samples. For FP galaxies, these could be used to investigate the impact of fiber position accuracy on velocity dispersion measurements. For TF galaxies, they could be used to build more accurate rotation curves to validate our large sample with fewer fibre positions. Accounting for fiber positions already included as part of the FP or TF selections, our ``EXT'' sample adds up to a total of 20 486 targets for 2 267 galaxies.

The remainder of the 2 911 extended galaxies consists of 644 SGA galaxies within the DESI footprint that also have semi-\textit{minor} axes greater than the fiber patrol radius; these objects, which we denote Extended Off-Axis (``EOA''), will always have a DESI fiber fall somewhere within the galaxy and not necessarily within reach of the existing extended object fiber positions. Hence, in addition to the targets in steps of $0.2 R_{26}$ along the semi-major axis, we also manually defined targets within the SGA ellipse but off the semi-major axis. For each of these galaxies, we undertook a small `citizen science' style project with help from undergraduates at the University of Rochester; selecting interesting locations by hand on which to place these fibers. These typically include \ion{H}{II} regions, small background galaxies, galaxy bars or warps, and interacting regions; avoiding obvious stars. In total, the 644 EOA galaxies are converted into 15 461 targets. A complete list of these targets is provided in electronic form as outlined in Table \ref{tab:target_ext}. 

\subsection{Target priorities and overview}

\begin{figure*}
 \includegraphics[width=0.8\textwidth]{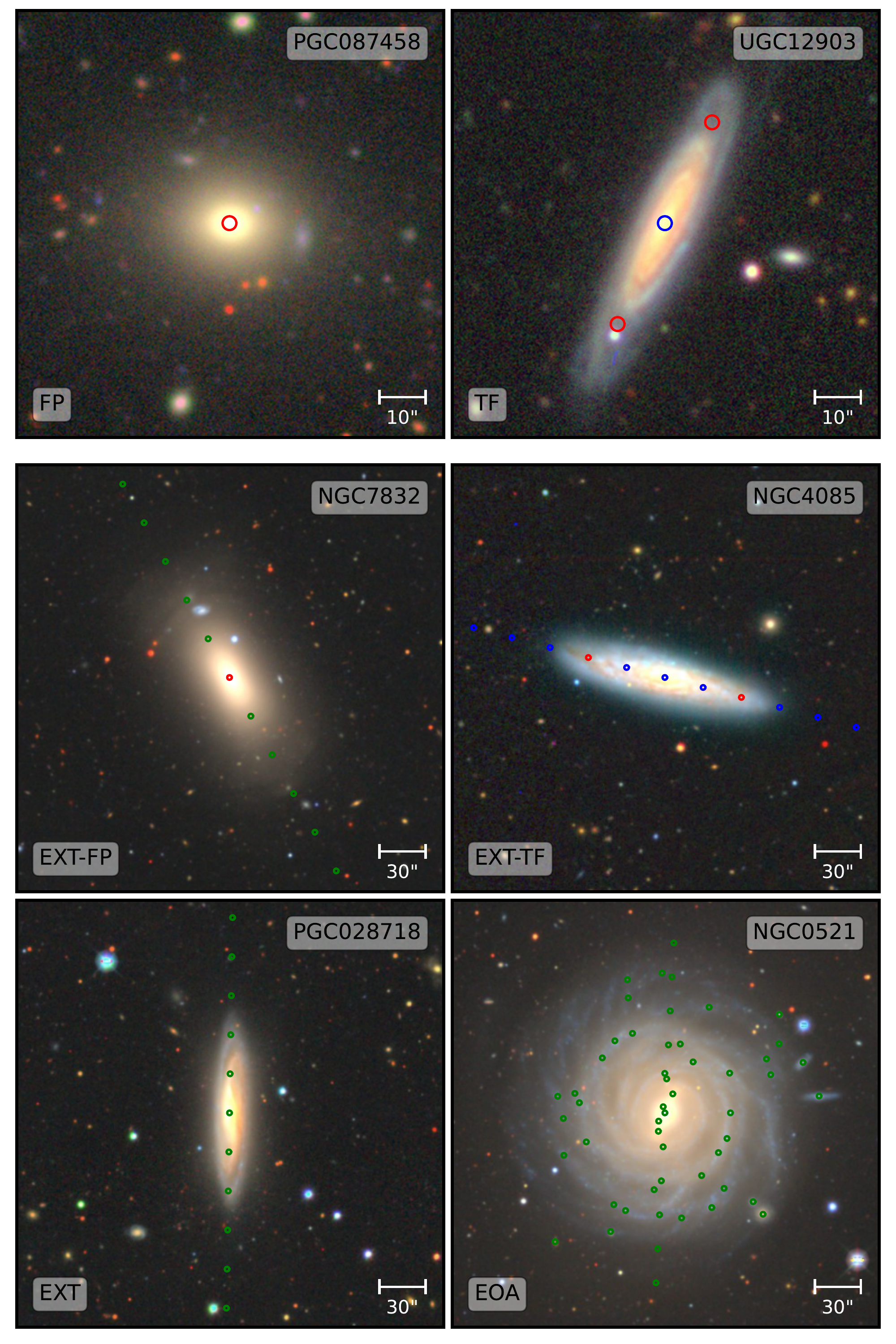}
 \caption{Postage stamps from the DESI Legacy Imaging Surveys DR9 showing example DESI PV galaxies and their corresponding targets. Different coloured circles represent different priority targets --- red, blue and green correspond to \texttt{HIGH}, \texttt{MEDIUM} and \texttt{LOW} priority respectively --- and the size of the circles is equal to the DESI fibre size. See Section~\ref{sec:methods} for a description of how targets are chosen for each galaxy class.}
 \label{fig:TS_example}
\end{figure*}

\begin{table}
    \centering
    \begin{tabular}{c|c|c|c|c}
    Sample & $N_{\mathrm{sample}}$ & Target Class & $N_{\mathrm{targets}}$ & Priority \\ \hline \hline
       FP  &  427 273 & - & 427 273 & \texttt{HIGH} \vspace{4pt} \\ 
       \multirow{2}{*}{TF} & \multirow{2}{*}{ 129 772 } & Centre &  129 772 & \texttt{MEDIUM} \\
         &  & On-axis at $\pm0.4 R_{26}$ & 259 544 & \texttt{HIGH} \vspace{4pt} \\ 
       \multirow{3}{*}{EXT} & 722 & Also in FP & 7 220 & \texttt{LOW} \\
         & 1 243 & Also in TF & 9 944 & \texttt{MEDIUM} \\
         & 302 & Not in FP or TF & 3 322 & \texttt{LOW} \vspace{4pt} \\ 
       SGA  &  81 611 & - & 81 611 & \texttt{LOW}\\
       EOA  & 644  & - & 15 461 & \texttt{LOW}\\ \hline
       Total & 638 959 & - & 934 147 & - \\
    \end{tabular}
    \caption{Numbers of galaxies and targets in the DESI PV subsamples, and their relative priority as related to their importance to our primary science goals. See Section~\ref{sec:sample} for a description of how these galaxy sample and their targets are defined. The total value of $N_{\mathrm{sample}}$ (the number of galaxies) is not quite the sum of all rows, because the EOA galaxies are wholly contained in the EXT sample, and subsets of the EXT galaxies are also FP or TF galaxies.}
    \label{tab:targets}
\end{table}

As mentioned in this Section, we have five samples of galaxies (FP, TF, SGA, EXT and EOA) which define a series of targets. These are, in turn, recombined into three different priority groups based on their relative importance to the goals of the PV survey. Targets with \texttt{HIGH} priority are deemed as essential for obtaining the PV measurements; \texttt{MEDIUM} priority are targets of interest that enable more careful study of how well we can measure the rotation speed of our TF sample; and \texttt{LOW} priority are targets that are less useful for the PV survey, but may be of more general scientific use or provide extragalactic targets in regions of the sky where DESI would otherwise not be able to place a science fibre. All targets are provided for both bright (when survey speed is between 10-40\%) and dark (survey speed above 40\%) time. The survey speed quantifies the observing conditions such as sky brightness (moon or twilight), sky transparency, and seeing. A summary of the numbers of objects and targets in each sample, as well are their relative priorities, is presented in Table~\ref{tab:targets}. In total, we have 934 147 unique targets spread across 638 959 galaxies. We have 686 817, 139 716, and  107 614 targets (corresponding to $73.5\%$, $15.0\%$, and $11.5\%$ of the total) in the \texttt{HIGH}, \texttt{MEDIUM} and \texttt{LOW} priority lists respectively. Examples of our target classes, and their respective fibre placements and priorities, are shown in Fig.~\ref{fig:TS_example}. The target priorities for each target are provided in Tables \ref{tab:target_fp} to \ref{tab:target_ext}. 

\section{Validating the target selections}
\label{sec:results}

In this section, we validate our two primary target selections for the FP and TF sample, demonstrate that our selections are targeting the correct types of objects, and show that we expect to be able to recover sufficient quality DESI spectra to produce PVs with these samples. In this section, we also quantify our expected ``success'' rate, i.e., how often we expect to be able to convert a target galaxy into a PV after the DESI observations have been made, and forecast numbers for our final PV sample. When deciding on our target selection, we had to strike a delicate balance between a sufficiently pure sample that does not waste too many fibres on targets that would not be useful for PV studies and avoiding cuts that could notably bias the distribution of our intended targets that would have to be considered in our models later. 

\subsection{Truth catalogues}
\label{sec:truth}
We created sets of truth catalogues that use previously established morphological classifications. Our two main sources are GalaxyZoo \citep{GalaxyZoo,GalaxyZoo_data} and SGA \citep{MoustakasSGA}. GalaxyZoo used SDSS images to run a Citizen science projects that had volunteers carry out the morphological classifications and the relative fraction of votes of these citizen scientists and additional corrections for biases was used to define the morphological types presented in their catalogues. The morphological classifications in SGA uses the HyperLeda\footnote{\url{http://leda.univ-lyon1.fr}} database that complies various sources from the literature. Hence their morphological classifications are limited to objects that were covered by previous surveys and observations. GalaxyZoo data are only available within the footprint of the SDSS main galaxy survey, and the SGA catalogue also contains a higher rate of incompleteness in its morphological classifications outside this, so we restrict our comparisons to the confines of this footprint. Using these restrictions, we are able to use these subsets of SGA and GalaxyZoo as truth catalogues for the validation of our target selection criteria. However, we found that our truth catalogues are not perfect as they disagree with each other at times (already visible by eye in the distribution of ETGs in Figure \ref{fig:colourcuts}) and visual inspection carried out by experts disagreed for a fraction of the objects as well. This is a common problem with classifications (see \citealt{Goode:2022}), as visual inspections are usually good at finding clearly wrong/bogus objects of a certain target class, but are less clear about the correct objects with a wide spectrum of ambiguity in between. Furthermore, the outcome of our tests also depends on how we used these truth catalogues --- in the case of GalaxyZoo the morphological classification comes with a probability, whereas in the case of the SGA catalogue the morphologies are split into very fine sub-types for which one could debate whether or not such galaxies should be considered part of our sample. If the classification is too restrictive, then only very few objects fall within it and we would miss out on many galaxies that we want in our sample.

We designed specific sets of truth catalogues for ETGs using our main sources (SGA and GalaxyZoo). In the case of the SGA, we defined three truth samples with different degrees of strictness: securely identified ellipticals (classification: \texttt{E} only), securely identified ETGs (classification: \texttt{E}, \texttt{S0}, and \texttt{E-S0}), and probable ETGs (classification: \texttt{E}, \texttt{S0}, \texttt{E-S0}, \texttt{DEV}, and \texttt{E?}). In the case of GalaxyZoo, we defined two main truth samples based on the debiased vote ratio \texttt{p\_el\_debiased} of galaxies being ellipticals of greater than 0.5 and 0.8. We also account for our cut in the axis-ratio in the truth catalogues. 

In similar fashion we create our truth catalogues for LTGs. Based on the SGA \citep{MoustakasSGA}, we define two truth samples with different degrees of strictness: one with securely identified LTG (classification: \texttt{SABa} to \texttt{SABd}, \texttt{SBa} to \texttt{SBd}, and \texttt{Sa}, to \texttt{Sd}) and the second also including probable LTG (classification: \texttt{SABa} to \texttt{SABd}, \texttt{SBa} to \texttt{SBd}, \texttt{Sa}, to \texttt{Sd}, \texttt{S?}, \texttt{.S?\ldots}, \texttt{.SB?\ldots}, and \texttt{Sm}). In the case of GalaxyZoo \citep{GalaxyZoo}, we define two main truth samples based on the debiased vote ratio \texttt{p\_sp\_debiased} greater than 0.5 and 0.8. Again, we also account for our cut in the axis-ratio in the truth catalogues.

\subsection{FP sample validation}
\label{sec:results_FP}

\begin{figure}
 \includegraphics[width=\columnwidth]{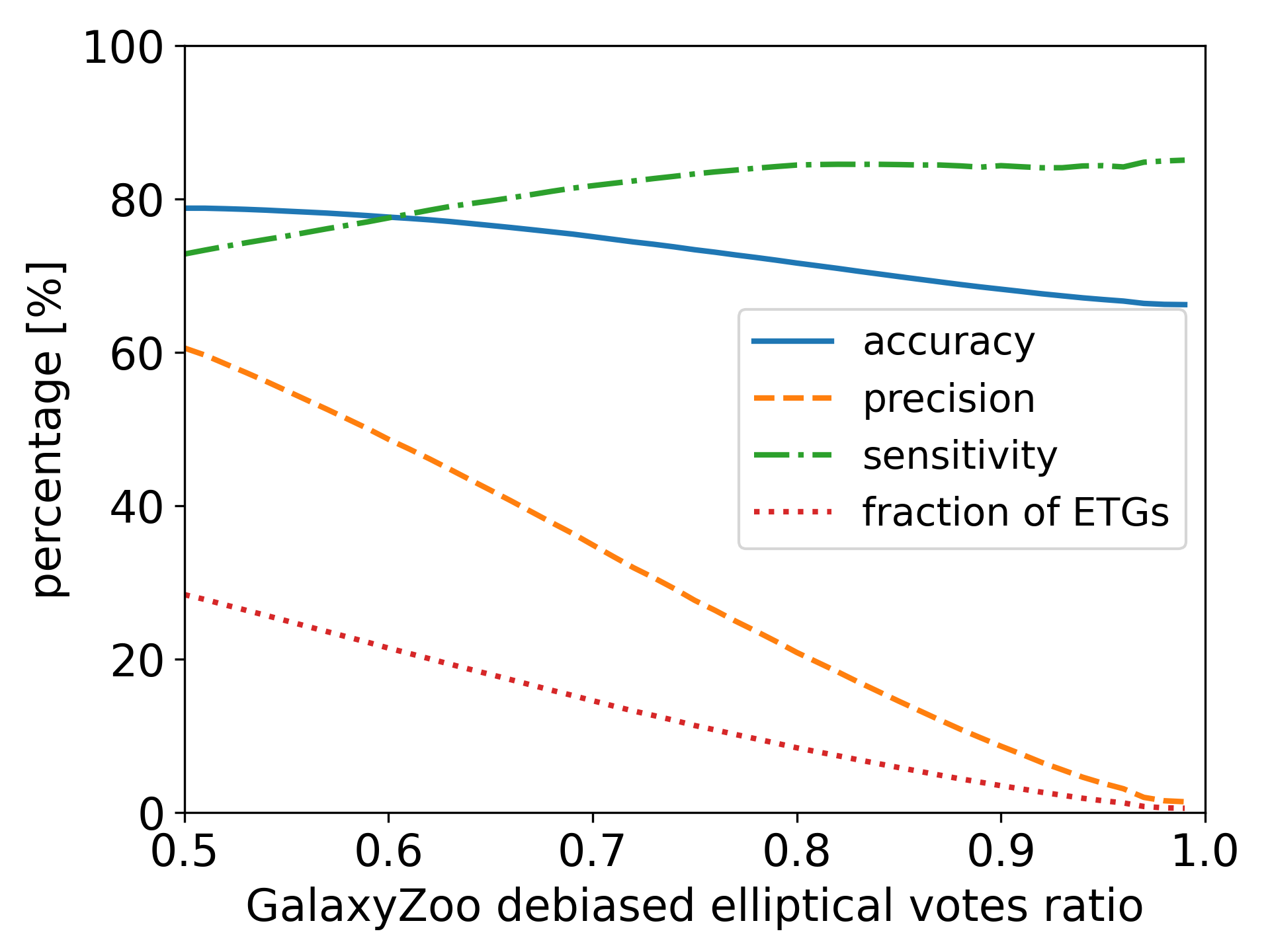}
 \caption{Quality evaluation parameters for our classification using elliptical galaxies in GalaxyZoo with different minimum probabilities as a truth sample.}
 \label{fig:galaxyzoo_quality_etg}
\end{figure}

\begin{table*}

\begin{center}
\begin{tabular}{c|cccccccccc}
Truth catalogue & TP & FN & FP & TN & accuracy & precision (T) & precision (F) & sensitivity & support (T) & support (F) \\
\hline \hline
SGA ellipticals & 23311 & 16396 & 25192 & 96899 & 0.743 & 0.481 & 0.855 & 0.587 & 39707 & 122091 \\
SGA secure ETGs & 26215 & 17566 & 22288 & 95729 & 0.754 & 0.54 & 0.845 & 0.599 & 43781 & 118017 \\
SGA probable ETGs & 27555 & 24313 & 20948 & 88982 & 0.72 & 0.568 & 0.785 & 0.531 & 51868 & 109930 \\
GalaxyZoo ETG (p>0.8) & 29009 & 5350 & 109913 & 262301 & 0.717 & 0.209 & 0.98 & 0.844 & 34359 & 372214 \\
GalaxyZoo ETG (p>0.5) & 84155 & 31403 & 54767 & 236248 & 0.788 & 0.606 & 0.883 & 0.728 & 115558 & 291015 \\
\end{tabular}
\caption{Quality statistics of our photometric selection criteria of ETGs for various truth catalogues. The columns are TP: true positives; FN: false negatives; FP: false positives; accuracy: accuracy value computed from the confusion matrix for the given truth catalogue; TN: true negatives; precision (T): precision value for the respective truth sample, precision (F): precision value for the respective false sample; support (T): number of objects in the truth sample; support (F): number of objects in the false sample. See text for details.} 
\label{tab:confusion_etg} 
\end{center}
\end{table*}

\begin{figure}
 \includegraphics[width=\columnwidth]{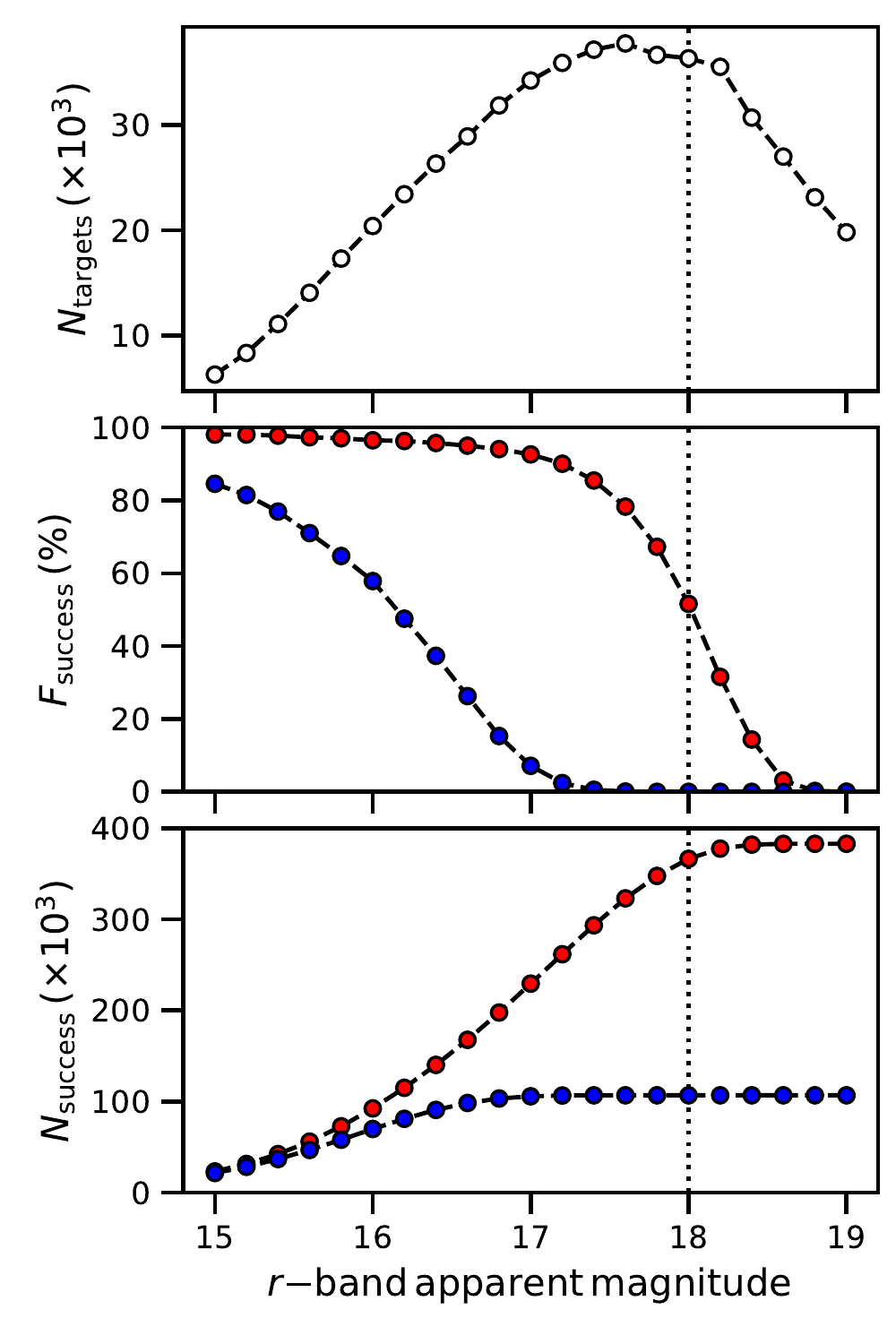}
 \caption{Histograms as a function of $r$ band magnitude, in bins of width 0.1 mag. \textit{Vertical dotted line:} magnitude limit of FP target selection. \textit{Top}: the number of FP targets in our sample. \textit{Middle}: $F_{\mathrm{success}}$; the percentage of targets we expect to obtain accurate ($<10$\% relative uncertainty) velocity dispersions for with a single DESI Bright or Dark time observation (blue and red points respectively). \textit{Bottom}: the cumulative number of FP targets we expect to obtain accurate velocity dispersions for in Bright and Dark time.}
 \label{fig:FPsims}
\end{figure}

\begin{figure*}
    \centering
    \includegraphics[width=0.99\textwidth]{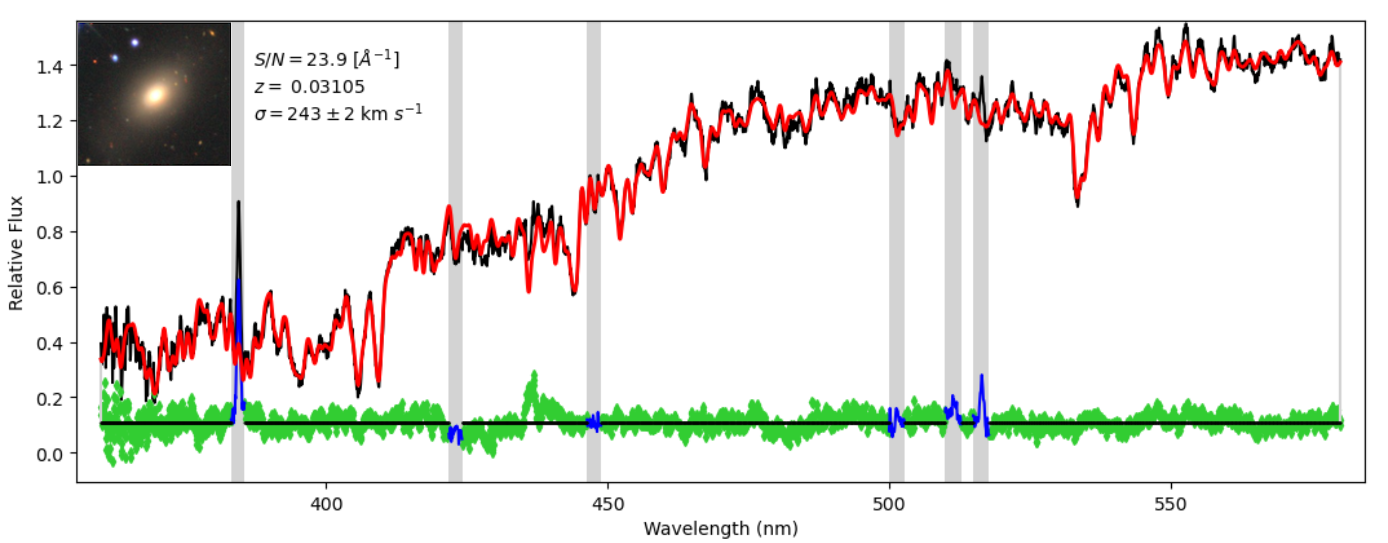}
    \caption{\texttt{pPXF} fit to a blue arm spectrum (target id = 39628414323262865, with Legacy Survey DR9 image shown in the top left corner). The black line is the relative flux of the observed DESI spectrum. The red line is the \texttt{pPXF} fit for the stellar component. The green symbols at the bottom are the fit residuals. The grey bands show masked gas emission lines, with best-fitting gas-only models given in blue. Signal-to-noise ratio, redshift, and velocity dispersion for this galaxy is printed in the top left corner. This spectrum is among the highest SNR ratio of the FP sample with SNR = 23.9 \AA$^{-1}$ and the typical SNR for the FP survey is 10 \AA$^{-1}$.}
    \label{fig:ppxf_fit}
\end{figure*}

We tuned the selection criteria for our FP sample given in Section~\ref{sec:ETG} by comparing our sample of ETGs to morphological classifications from truth catalogues based on GalaxyZoo and SGA. When designing the colour-cuts, we used the most strict classifications as guidance for the region of parameter space that we definitely wanted to include, and then less strict classifications with additional visual inspection to fine-tune the boundaries, thereby allowing for reasonably large and unbiased samples of targets. 

As a more solid assessment of the quality of our classification, we calculated the confusion matrix for these truth samples and derived various evaluation parameters for them. The results are listed in Table \ref{tab:confusion_etg}. Additionally, we provide the SGA classifications and GalaxyZoo classifications used for our tests in Tables \ref{tab:sga_targets} and \ref{tab:zoo_targets}, respectively. We use the established measures and terms, such as: support of truth (T) or false (F) sample, which are the numbers of objects in the respective categories according to the truth catalogue. The true positive (TP) sample are the objects within the truth sample that are correctly predicted by the model. Those that are within the truth sample, but are missed by the model prediction are called false negatives (FN). There are also objects are predicted are incorrectly predicted to be in the truth sample by the model, which are called false positives (FP). The objects that are correctly predicted to be in the false sample are refereed to as false negatives (FN). We can further define the accuracy as (TP+TN)/(TP+TN+FN+FP), the sensitivity as TP/(TP+FN), the precision of the truth sample as TP/(TP+FP), and the precision of the false sample as TN/(TN+FN). Overall our classification achieves consistently above $70\%$ accuracy and also high values of sensitivity for the GalaxyZoo samples (a little less so for SGA). The precision for the truth sample strongly depends on the support of truth (see Figure \ref{fig:galaxyzoo_quality_etg} in which we show the dependence of diagnostic parameters on the GalaxyZoo classification ratios), hence when we restrict our truth sample too much, we record a large fraction of false positives that in reality are actually true positives for galaxies lying on the FP. Figure \ref{fig:galaxyzoo_quality_etg} also shows that accuracy and sensitivity are only influenced little by the strictness of our truth catalogues.  

Based on our tests using the various truth catalogues, we can pessimistically estimate the success rate of our FP target selection to be about $50\%$. However, factoring in additional visual inspection of false positives and preliminary results from early spectroscopic observations, the success rate might actually be closer to two-thirds of our targets. More precise estimates of the success rate will be provided in the upcoming paper (Said et. al., in prep.) on the FP calibration using DESI survey validation data.

In addition to testing the quality of our colour-cuts, we also tested the magnitude limit in our selection. Our imposed limit of $r < 18$, that we used in the selection of the FP sample, is a full magnitude deeper than the SDSS sample used in \citet{Howlett2022}, and is designed to remove galaxies that would have very low chance of a successful velocity dispersion measurement. The exact value for this cut was decided by simulating the signal-to-noise ratios of our early-type targets using the DESI $\mathtt{specsim}$ package\footnote{\href{https://specsim.readthedocs.io}{https://specsim.readthedocs.io}}, which takes into account the DESI instrument characteristics \citep{Guy2022}. We then set a threshold of signal-to-noise ratio (SNR) of $>7.5$\AA$^{-1}$ after a single 1000s dark time exposure in the B-band arm of the DESI spectrograph as the threshold for obtaining a usable velocity dispersion with relative error of less than 10\%, and evaluated the fraction of galaxies that we would expect to pass this threshold in different magnitude bins.

This fraction is shown in Fig.~\ref{fig:FPsims}, alongside the number of targets in each 0.1 magnitude bin. At a limit of $r=18$, the simulated SNR is high enough that $\sim50\%$ of our FP targets are predicted to have a successfully measured velocity dispersion after a single dark-time exposure. There is little benefit to targeting galaxies fainter than this, both because there are not many of them given we already apply cuts on photometric redshift and because the SNR for these galaxies drops rapidly. This can be seen in the bottom panel of Fig.~\ref{fig:FPsims} where we plot the cumulative number of FP observations that we simulate as having sufficient SNR to measure the velocity dispersion. Beyond $r=18$ the cumulative number quickly plateaus. As a final note, it is worth mentioning that although $\sim50\%$ may seem low, because our targets use spare fibres it is possible that a given galaxy can obtain multiple dark time exposures, especially if they are on different tiles. Said et. al., (in prep.) explore the properties of such `repeat' DESI observations in more detail, but at $r=18$ the fraction of objects that pass our SNR threshold increases to $\sim85\%$ after two repeats. As an example of a successful observation of a FP target, we provide the \texttt{pPXF} \citep{pPXF1,pPXF2} fit of one such object in Figure \ref{fig:ppxf_fit}. It illustrates that we can clearly recover absorption lines and their shapes from dark time spectra.

\subsection{TF sample validation}
\label{sec:results_TF}

\begin{figure}
 \includegraphics[width=\columnwidth]{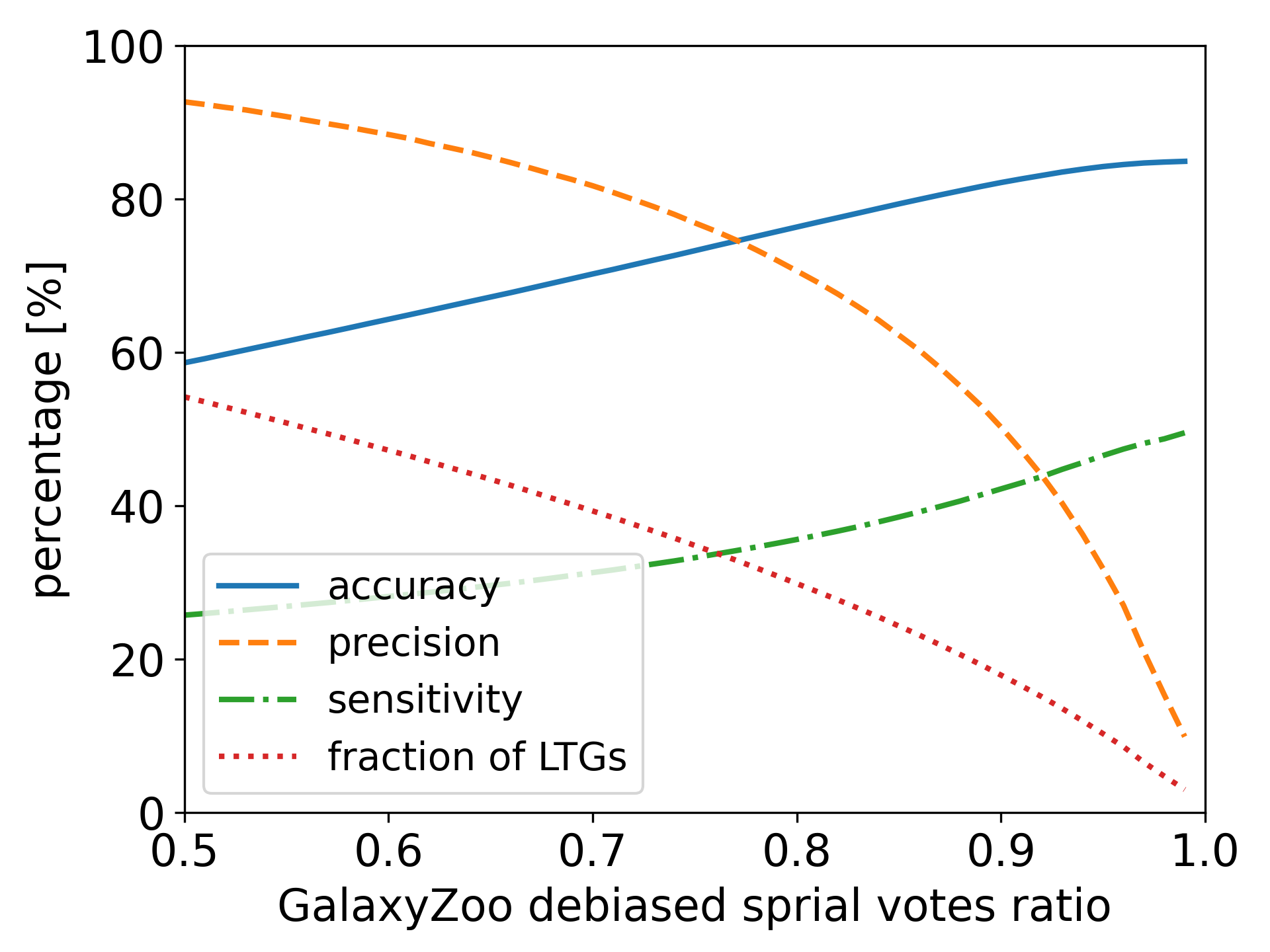}
 \caption{Quality evaluation parameters for our classification using spiral galaxies in GalaxyZoo with different minimum probabilities as a truth sample.}
 \label{fig:galaxyzoo_quality_ltg}
\end{figure}

\begin{table*}
\begin{center}
\begin{tabular}{c|cccccccccc}
Truth catalogue & TP & FN & FP & TN & accuracy & precision (T) & precision (F) & sensitivity & support (T) & support (F) \\
\hline \hline
SGA secure LTG & 38304 & 26372 & 27949 & 69173 & 0.664 & 0.578 & 0.724 & 0.592 & 64676 & 97122 \\
SGA probable LTG & 56390 & 44321 & 9863 & 51224 & 0.665 & 0.851 & 0.536 & 0.56 & 100711 & 61087 \\
GalaxyZoo LTG (p>0.8) & 43234 & 78169 & 17988 & 267182 & 0.763 & 0.706 & 0.774 & 0.356 & 121403 & 285170 \\
GalaxyZoo LTG (p>0.5) & 56735 & 163630 & 4487 & 181721 & 0.587 & 0.927 & 0.526 & 0.257 & 220365 & 186208 \\
\end{tabular}
\caption{Quality statistics of our photometric selection criteria of LTG for various truth catalogues. Column names are as described in Table~\ref{tab:confusion_etg}.} 
\label{tab:confusion_ltg} 
\end{center}
\end{table*}

\begin{figure*}
    \centering
    \includegraphics[width=\textwidth]{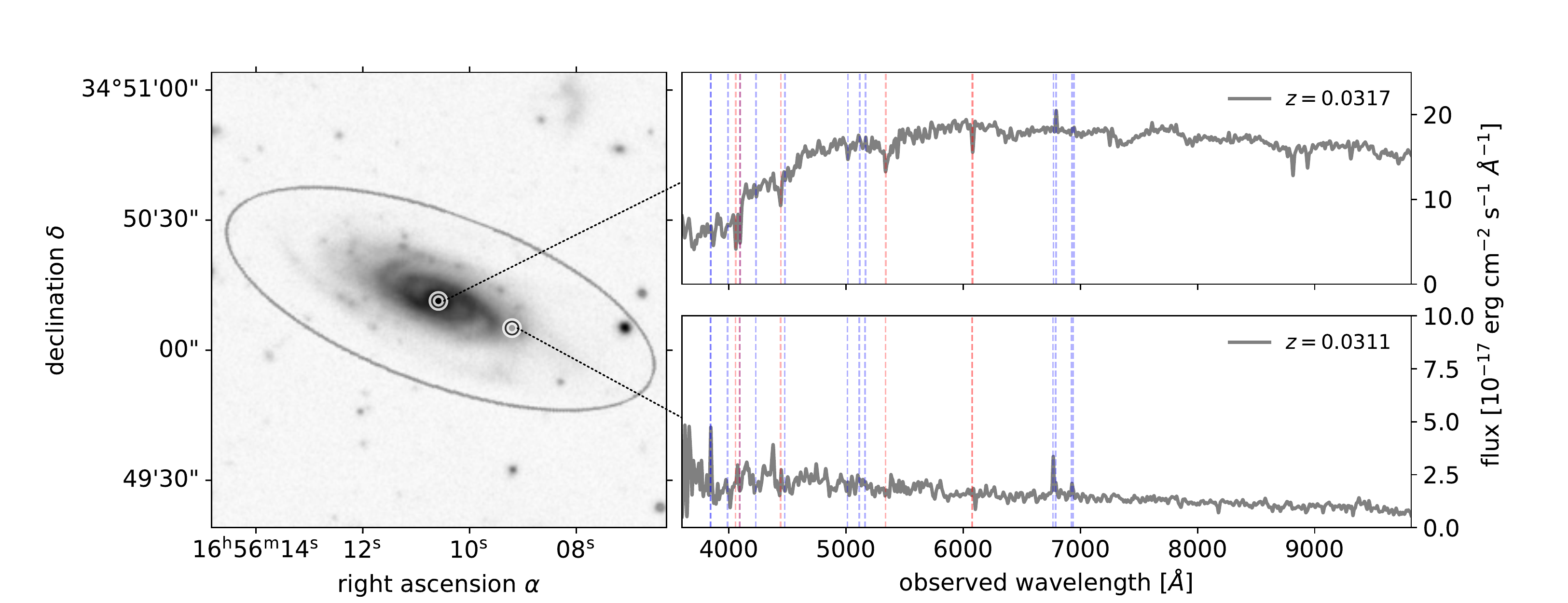}
    \caption{Left: DESI Legacy Imaging Survey DR9 measurement of the spiral galaxy UGC~10615 
    showing $R_{26}$, the 26 mag~arcsec$^{-2}$ isophote (solid line), and the positions of two early DESI spectroscopic observations at the galactic nucleus and semimajor axis at $0.33 R_{26}$ (the final targeting criteria uses $0.4 R_{26}$). Right: corresponding observed spectra, with major emission and absorption lines indicated in red and blue, respectively.}
    \label{fig:desi_obs_rotational}
\end{figure*}


\begin{figure*}
   \centering
    \includegraphics[width=\textwidth]{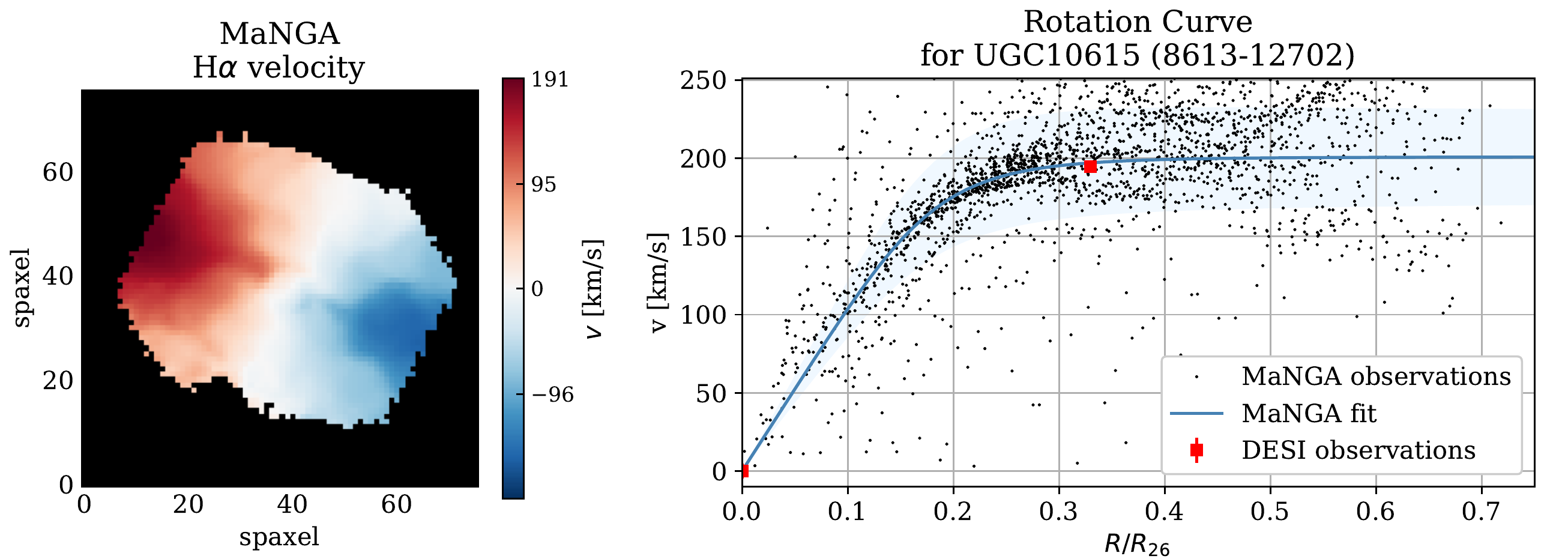}
    \caption{Comparison between a DESI rotational velocity measurement and SDSS MaNGA observations of the same galaxy.  \emph{Left:} 
    SDSS MaNGA DR15 H$\alpha$ velocity map.  
    \emph{Right:} Deprojected SDSS MaNGA velocities (black points) and corresponding best-fit rotation curve \citep[blue curve;][]{Douglass22}.  The rotational velocity measured by DESI is shown in red and agrees well with the SDSS MaNGA results.}
    \label{fig:MaNGA_comp}
\end{figure*}

We test the photometric target catalogues for the late-type galaxies suitable for the TF relation using the truth catalogues create for them. Following the same procedure as for the ETG, we calculate the confusion matrix and evaluation parameters for our LTG classifications; the results are listed in Table \ref{tab:confusion_ltg}. We found that the overall accuracy is around or above $60\%$. Additionally, the precision for more generous classifications of the truth sample is remarkably high ($>92\%$), which indicates the objects we have photometrically selected mostly consist of spiral galaxies (of some sub-type) with few false positives. This trend is also highlighted when we study a range of truth samples with varying degrees of strictness in their classifications using GalaxyZoo as shown in Figure \ref{fig:galaxyzoo_quality_ltg}. These results allow us to expect an up to $85\%$ successful photometric identification of late-type galaxies. 

In addition to confirming our successful targeting of TF-relation targets, we also confirm that we successfully collect the required observational data. To measure the rotational velocity of a galaxy at a given radius, we measure the difference in redshift between the galaxy center (its systemic velocity, due to a combination of the expansion of the Universe and the PV of the galaxy) and at a point along the galaxy's semi-major axis.  An example of this is shown in Fig.~\ref{fig:desi_obs_rotational}.  Since the TF relation is relating a galaxy's luminosity and mass (via the rotational velocity), our goal is to measure the rotational velocity at a distance sufficiently far from the galaxy center so as to obtain the maximum velocity of the galaxy's rotation curve. This can be difficult at the wavelengths of visible light, since the S/N of the galaxy's spectrum is often very low at the galaxy's outskirts.  To determine how far from the galaxy center DESI is able to robustly determine the redshift, we placed fibers at $0.33 R_{26}$, $0.67 R_{26}$, and $R_{26}$ along the semimajor axis of each galaxy, in addition to the galaxy center, during survey validation (SV3).

We find that DESI is able to robustly measure the redshift in 91.7\% of the spectra observed at $0.33 R_{26}$, while only 28.3\% of the spectra observed at $0.66 R_{26}$ are successful (and only 12.8\% are successful at $R_{26}$). Combined with the results of \cite{Schlegel_PhD}, we choose to measure the rotational velocity at $0.4 R_{26}$ during the DESI main survey. 

To assess the success of converting our measured redshifts to the asymptotic rotational velocities of the late-type galaxies with DESI, we compare our SV3 measurements to observations made of the same galaxies in the SDSS MaNGA DR~15 survey \citep{Manga}. MaNGA is an integral field unit spectroscopy survey carried out within SDSS~IV. It observed $\sim$10 000 galaxies to study their detailed composition and kinematic structure. This wealth of information can be used for consistency checks of our fibre placements and observations. An example of one of these galaxies is shown in Figure~\ref{fig:MaNGA_comp}, where it is readily apparent that our DESI rotational velocity measurement agrees with the IFU data observed by MaNGA and recovers the maximum rotational velocity.

\begin{figure*}
    \centering
    \includegraphics[width=0.49\textwidth]{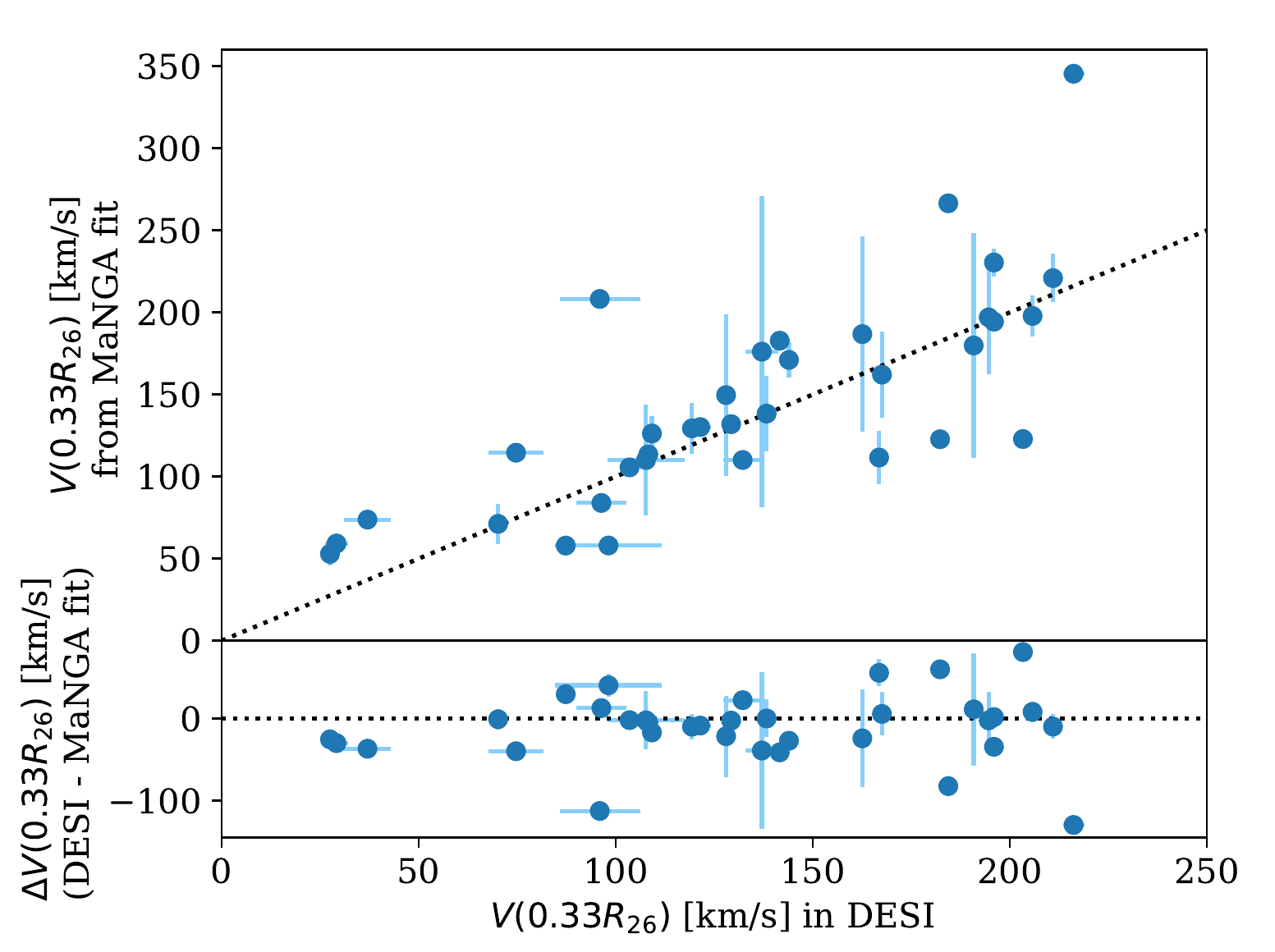}
    \includegraphics[width=0.49\textwidth]{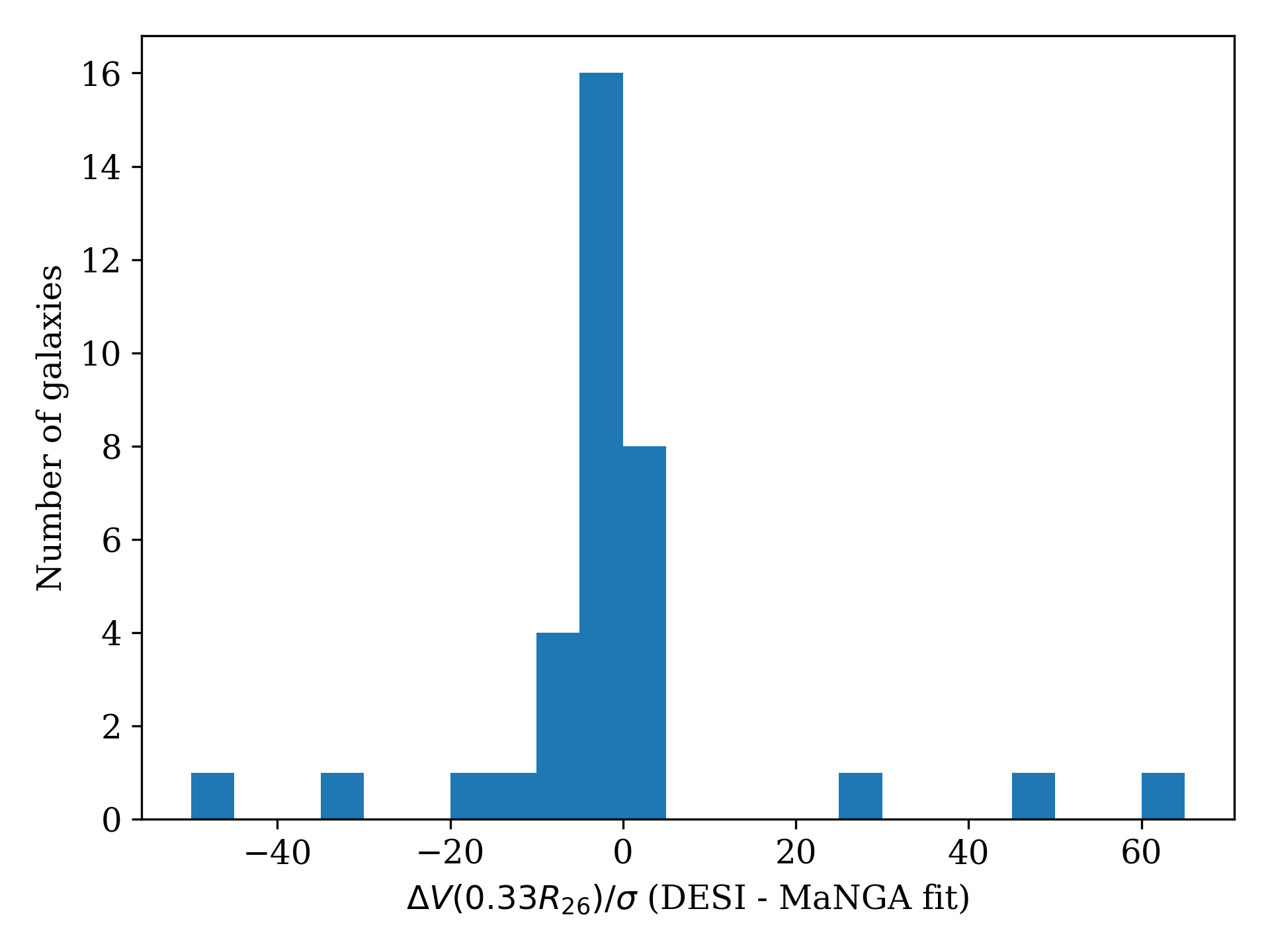}
    \caption{Comparison between the observed rotational velocities at $0.33 R_{26}$ and that calculated at the same radius from modeled rotation curves to the SDSS~MaNGA~DR15 H$\alpha$ velocity maps \citep{Douglass22} for those galaxies which overlap these two surveys.  The black dotted line on the left represents $y = x$; the distribution of the residuals normalized by their relative uncertainties is shown on the right.  
    Our targeting strategy in the DESI PV Survey successfully recovers the expected rotational velocity at this radius.}
    \label{fig:V0p33_comp}
\end{figure*}

\begin{figure*}
    \centering
    \includegraphics[width=0.49\textwidth]{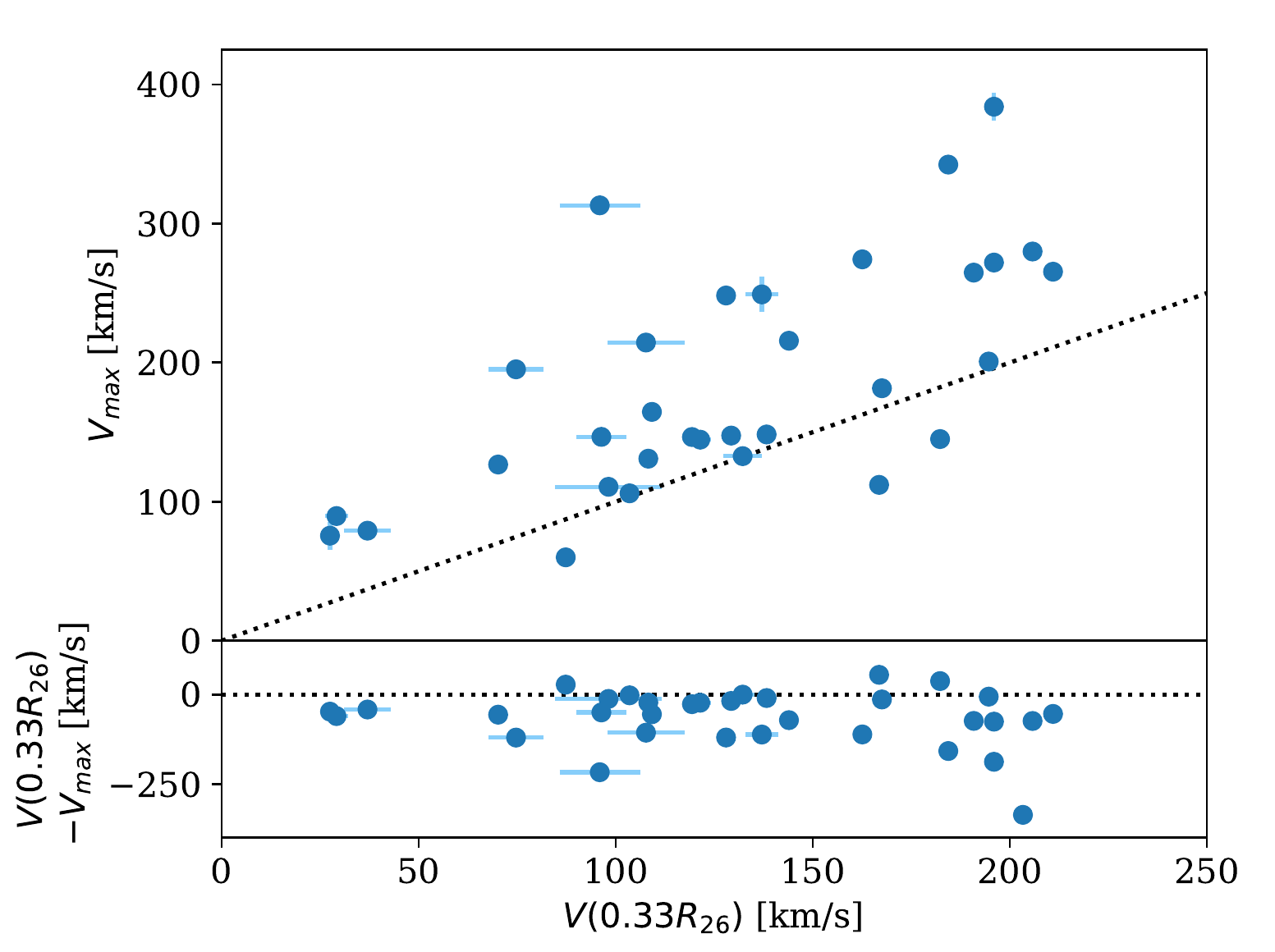}
    \includegraphics[width=0.49\textwidth]{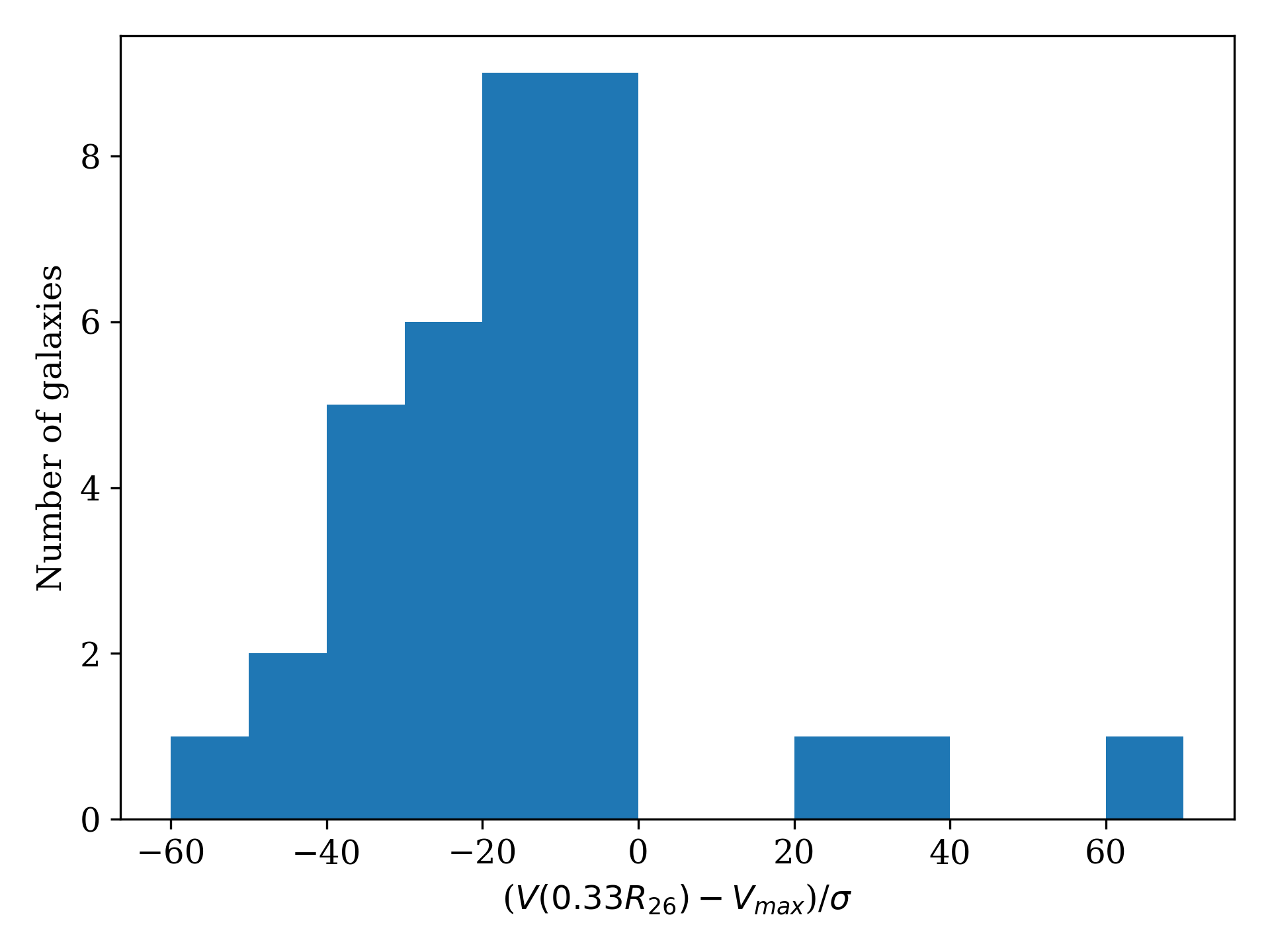}
    \caption{Comparison between the observed rotational velocities at $0.33 R_{26}$ and the fitted $V_\text{max}$ to the SDSS~MaNGA~DR15 H$\alpha$ velocity maps \citep{Douglass22} for those galaxies which overlap these two surveys.  The black dotted line on the left represents $y = x$; the distribution of the residuals normalized by their relative uncertainties is shown on the right.  The velocities at $0.33 R_{26}$ come close to, but are slight underestimates of, $V_\text{max}$.}
    \label{fig:Vmax_comp}
\end{figure*}

We observed 34 galaxies during SV3 that were also observed as part of MaNGA~DR15 and have valid rotation curve models from \cite{Douglass22}.  With this set of galaxies, we can confirm that our targeting strategy recovers the expected velocity at $0.33 R_{26}$, and that we observe the maximum rotational velocity.  As shown in Fig.~\ref{fig:V0p33_comp}, we find good agreement with the velocities expected at this orbital radius from the modeled MaNGA rotation curves.  When we compare our observed velocities at $0.33 R_{26}$ to the expected maximum rotational velocity, $V_\text{max}$, fit to the MaNGA data, Fig.~\ref{fig:Vmax_comp} shows that we either recover ($\sim$50\%) or slightly underestimate $V_\text{max}$. We expect that this bias will be reduced with our observations at $0.4 R_{26}$ because it is at a larger galactocentric radius than $0.33 R_{26}$, allowing us to better recover the maximum rotational velocity during the main DESI survey. Even with this bias, \cite{Yegorova07} show that the TF relation can be calibrated with rotational velocities consistently measured at a given galactocentric radius, although the scatter in the relation is reduced when the velocities approach the asymptotic rotational velocity. Our recovery of the maximum rotational velocity will be elaborated further in the upcoming paper on the TF relation in early DESI data (Douglass et. al, in prep.).

An additional source of uncertainty in our measurement of the rotational velocity is the placement of the fibers along the semimajor axis of the galaxy, which depends on the rotation angle of the galaxy (angle east of north of the semimajor axis). When comparing our results to MaNGA, we find that the photometric rotation angle reported in the SGA is not always aligned with the kinematic rotation angle. In theory, this can be corrected for by comparing the velocities measured along the semimajor axis of the galaxy with those along the semiminor axis.  For a given orbital radius, the offset between the photometric and kinematic rotation angles is equal to
\begin{equation}\label{eqn:phi_offset}
    \Delta \phi = \arctan \left( \left| \frac{V_\text{obs, semiminor}}{V_\text{obs, semimajor}} \right| \right).
\end{equation}
However, observations along the semiminor axis are not available for most of our galaxies because of their spatial extent, so this correction cannot be widely applied. As discussed in Douglass et al. (in prep.), we can use the MaNGA~DR15 observations to help quantify this systematic effect. They show that the offset between position angle given in the SGA and the kinematic rotation angle results in a measured rotational velocity that is $\sim$25~km/s larger on average; this difference is amplified for galaxies that are more edge-on.
 
\section{Survey characteristics}
\label{sec:survey}

\begin{figure*}
    \centering
    \includegraphics[width=0.99\textwidth]{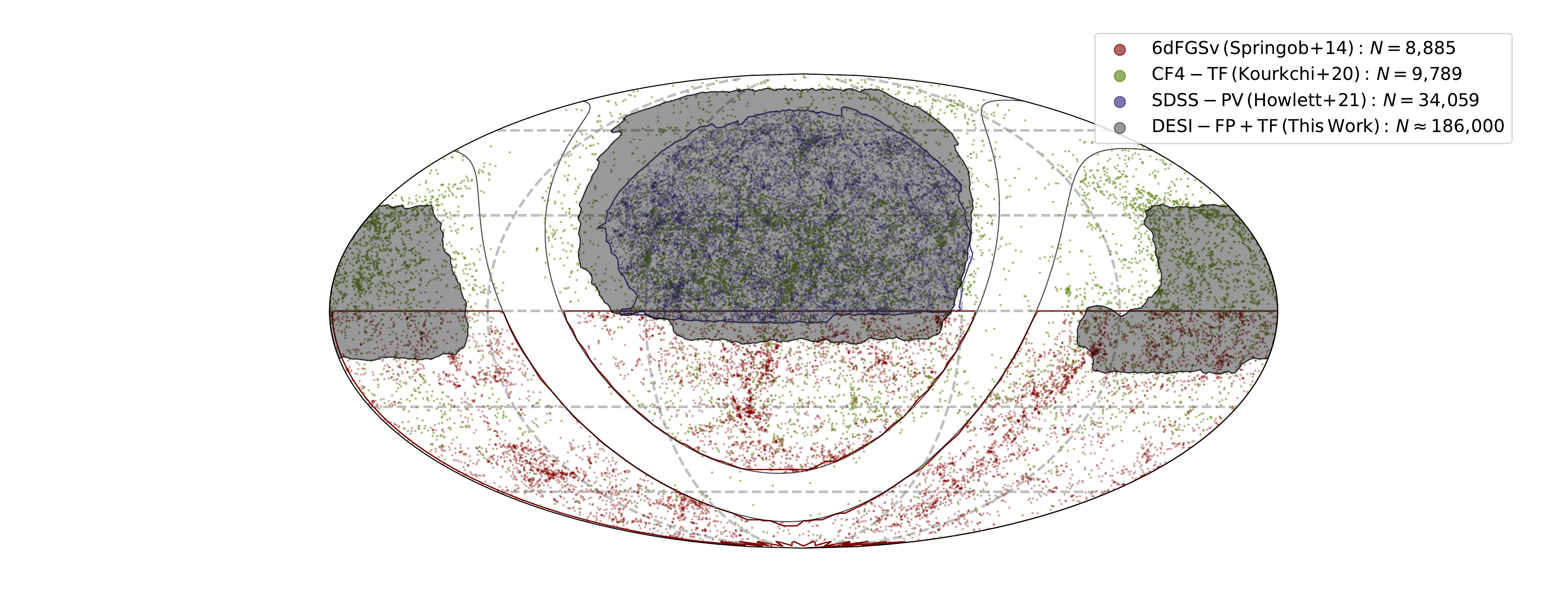}
    \caption{The expected footprint on the sky of the DESI-PV survey (gray shaded region), relative to existing PV data from the Cosmicflows-IV TF survey (CF4-TF; green points; \citealt{Kourkchi:2020}, the 6dFGSv survey (red points; \citealt{Springob2014}) and the SDSS-PV survey (blue points; \citealt{Howlett2022}).}
 \label{fig:DESIsky}
\end{figure*}

\begin{figure}
    \centering
   \includegraphics[width=0.48\textwidth]{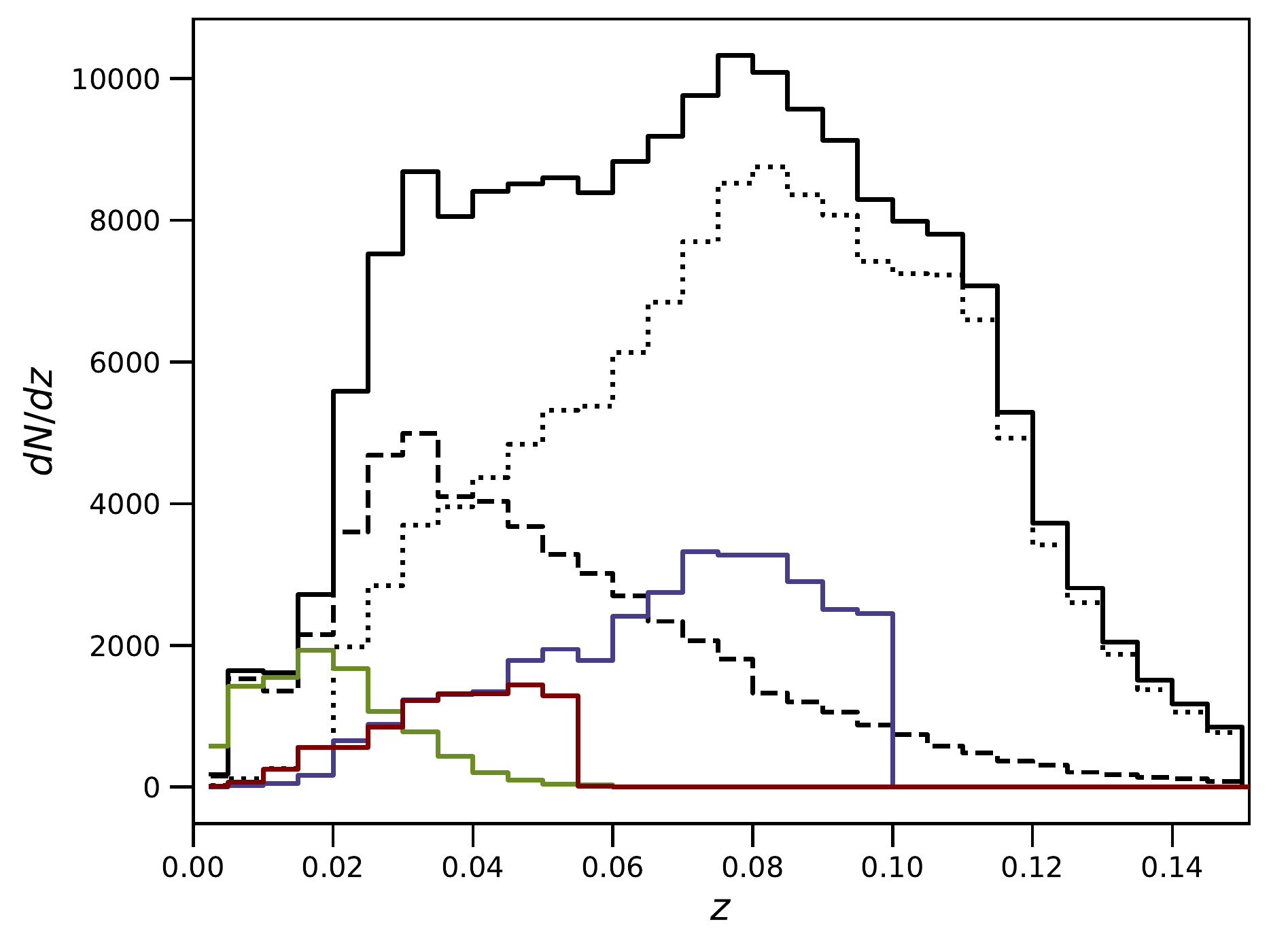} \\
   \includegraphics[width=0.48\textwidth]{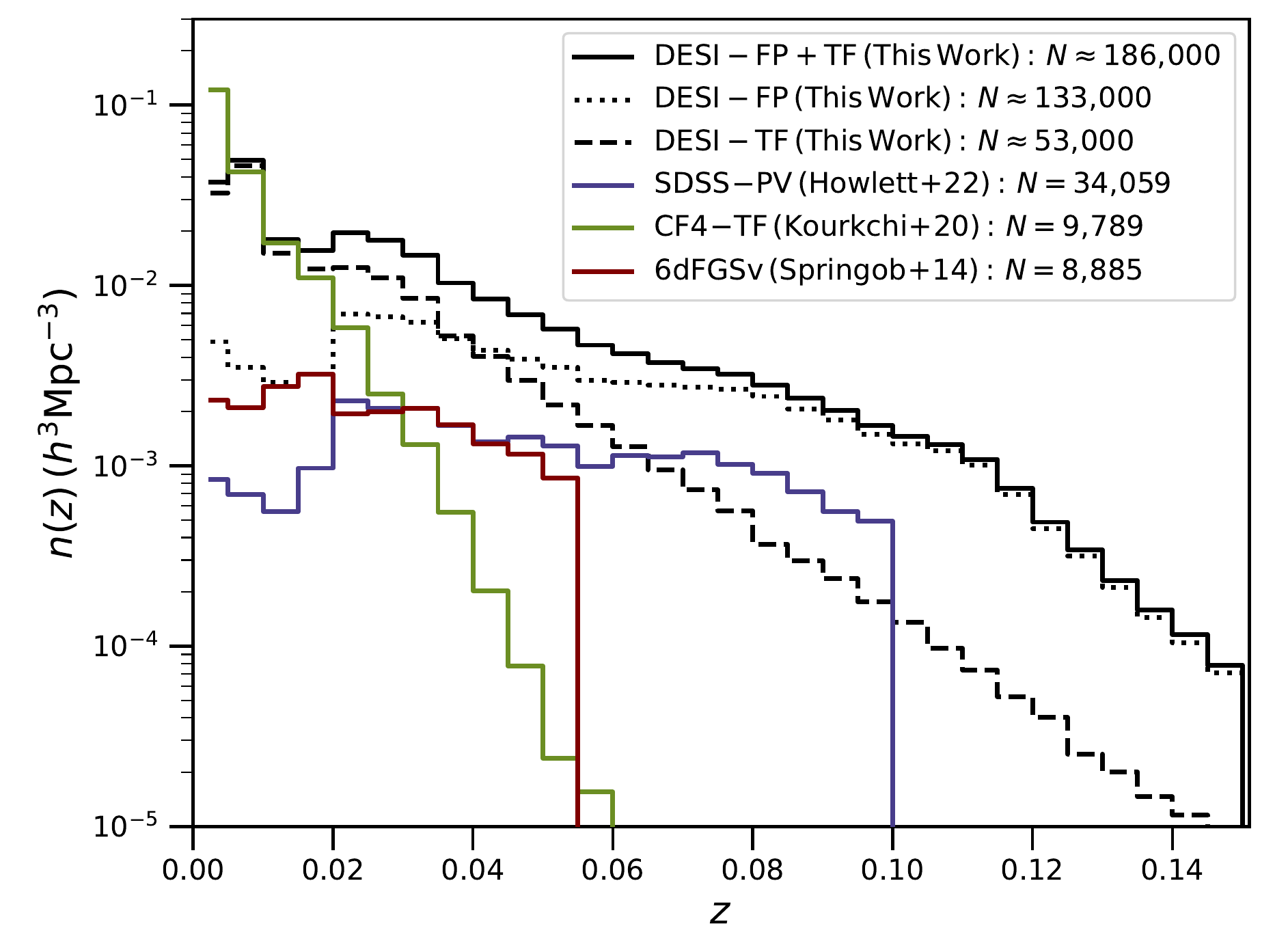}
   \caption{The photometric redshift distribution of the DESI PV survey including success probabilities for all of our targets. The top panel shows the number per 0.005 redshift bin, while the bottom panel is the number density per unit volume for our fiducial cosmology. In both cases the black solid, dashed and dotted lines correspond to the full, TF and FP DESI samples respectively. The green, red and blue lines are the distributions for the currently available Cosmicflows-IV TF (CF4-TF; \citealt{Kourkchi:2020}), 6dFGSv \citep{Springob2014} and SDSS-PV \citep{Howlett2022} surveys respectively.}
 \label{fig:DESInz}
\end{figure}

\begin{table*}
    \centering
    \begin{tabular}{c|c|c|c|c|c|c}
    \multirow{2}{*}{Sample} & \multirow{2}{*}{$N_{\mathrm{sample}}$} & \multirow{2}{*}{$N_{\mathrm{pointings}}$} & \multicolumn{3}{c}{$N_{\mathrm{allocated}}$} & \multirow{2}{*}{$N_{\mathrm{PV}}$} \\
     &  &  & Bright & Dark & Any & \\ \hline \hline
   FP  &  427 273 & $1$ & 288 123 & 118 091 & \textbf{307 722} & \textbf{132 824} \vspace{4pt} \\ 
   \multirow{3}{*}{TF} & \multirow{3}{*}{ 129 772 } & 1 &  59 526 & 30 947 & 42 970 & \multirow{3}{*}{ \textbf{53 513} } \\
    &  & 2 & 26 326 & 13 242 & \textbf{35 561} &  \\
    &  & 3 & 10 386 & 493 & \textbf{23 918} & \\
    \end{tabular}
    \caption{Numbers of all our $N_{\mathrm{sample}}$ FP and TF galaxies within the DESI footprint that are predicted to be assigned fibres during the full survey. $N_{\mathrm{pointings}}$ denotes the number of unique targets on each galaxy that receive at least one observation, and $N_{\mathrm{allocated}}$ is the number of galaxies that receive $N_{\mathrm{pointings}}$ observations under different conditions. The columns Bright and Dark are restricted to considering $N_{\mathrm{pointings}}$ \textit{only} under the stated conditions, whereas ``Any" is $N_{\mathrm{pointings}}$ under any conditions. $N_{\mathrm{PV}}$ is the number of successfully observed galaxies we expect to translate into PV measurements. In all cases, bold numbers indicate those that are actually usable data for our science (i.e., combining duplicates into a single redshift, and ignoring TF galaxies with only a single measurement).}
    \label{tab:fa_rates}
\end{table*}

Having validated our photometric selection and survey design with individual objects, in this section we present our predictions for what the the DESI PV survey is expected to yield by the end of the 5 years of operations, using simulations of the fiber-assignment process applied to the full DESI target list (including both main and secondary targets), and folding in our estimated PV success rate for the TF and FP selections from Section~\ref{sec:results}. The use of the fiber-assignment simulations allows us to account for the fact that, in general, our PV targets are included at lower priority than the main BGS targets and so are less likely to be observed. However, it is important to note that our FP targets and TF galaxy centres are all included in the BGS and so benefit from the higher bright-time priorities given to BGS in the ``merged'' target list. Accordingly, these targets achieve high fiber-allocations during bright time, and are only really ``spare-fiber'' targets during dark time. It is also important to note that the simulation represents only a single possible instance of the fiber-assignment process, and so while we expect this to be representative on average, the exact numbers of successfully observed targets and their redshift/angular distributions may differ from reality, and we cannot yet say whether or not any \textit{particular} galaxy will end up in our final sample.

Table~\ref{tab:fa_rates} presents the number of FP and TF galaxies that receive sufficient DESI fibers in the simulation such that we would have the necessary spectra and spectral S/N to attempt to measure a PV (which we denote $N_{\mathrm{allocated}}$). For the FP sample this means they received at least one observation in either bright or dark time. For the TF targets, we identify galaxies that were allocated up to three fiber observations at the centre and/or $\pm 0.4 R_{26}$ during bright and dark time. The respective bright and dark columns indicate the number of TF galaxies that received $N_{\mathrm{pointings}}$ \textit{only} under those conditions, but for the purposes of the PV sample we are able to relax this and use any mix of bright and dark observations (column ``Any'' in the table) to obtain observations at at least two distinct points on the galaxy. The column ``Any'' cannot be treated as the sum of the bright and dark columns because 1), there are some of the same targets that obtain both bright and dark observations (in which case we would use a weighted co-added spectrum with contributions from both exposures) and 2), because galaxies that obtain one pointing in bright time and a second different pointing in dark time would be entered, quite rightly, in both the $N_{\mathrm{pointings}} = 1$ rows for these two columns, but the $N_{\mathrm{pointings}} = 2$ row for "Any", reflecting the fact that the galaxy could now be used the TF analysis if we are agnostic to the conditions under which the observations were made.

The predicted number of galaxies allocated fibres is then multiplied by the success fraction for the different types of targets from Section~\ref{sec:results} to obtain the expected number of PV measurements we will obtain,
\begin{equation}
N_{\mathrm{PV}} = N_{\mathrm{allocated}} \times F_{\mathrm{success}}.
\end{equation}
For the TF sample, we assume a uniform $90\%$ success fraction, as justified in Section~\ref{sec:results_TF}. For the FP sample, we first use the estimated success fraction as a function of $r$-band apparent magnitude, given for single bright and dark time observations in Fig.~\ref{fig:FPsims}. As the fiber assignment simulations also include whether the target is observed in bright or dark time, we multiply each of our fiber-allocated input targets by the appropriate probability of them obtaining sufficient signal-to-noise for a velocity dispersion measurement.  We find that, overall, we retain a total of $\sim$58\% successfully observed targets compared to $N_{\mathrm{allocated}}$. This number, based on spectral SNR simulations, is similar to the number kept in the real SV data ($\sim$64\%; Said et al., in prep.) after applying a cut on the velocity dispersion uncertainty (which is what we actually prefer to cut on once measurements are available, but is heavily correlated with spectral SNR). This gives confidence that the spectral simulations are appropriate to use for our target validation. 

Based on this SV work, we then include an additional factor of $0.75$ to account for objects that would additionally be removed due to, for instance, the presence of H$\alpha$ emission in their spectrum or as a result of visual inspection. Overall, this amounts to the retention of $\sim$44\% of the FP targets in our final sample. We note that this could be increased somewhat in practice if any of our objects have repeat spectra in bright time that can be co-added to improve the SNR (which we have not considered in these predictions, but do occur in SV data). A reduction in the number of FP galaxies of $\sim$50--60\% is also fully consistent with that found in the SDSS PV catalogue of \cite{Howlett2022}.

Overall, our predicted DESI-PV sample consists of $\sim$133 000 successfully observed FP galaxies and $\sim$53 000 TF galaxies. The angular and redshift distributions of these samples (using the photometric redshifts from the input catalogues) are shown in Figs.~\ref{fig:DESIsky} and \ref{fig:DESInz}, respectively. Alongside these we show the distribution for some of the largest currently available PV catalogues; the Cosmicflows-IV TF sample at low redshift \citep{Kourkchi:2020}, which covers more or less the full sky; the southern hemisphere 6dFGSv survey \citep{Springob2014}; and the smaller area, but much deeper and more complete SDSS PV catalogue \citep{Howlett2022}. As can be seen in these figures, the DESI PV survey will provide the largest and deepest TF and FP samples to date, and cover an area of $\sim$14 000 deg$^2$, twice as large as the SDSS PV sample and similar to that of 6dFGSv.

\section{Cosmological forecasts}
\label{sec:coscast}

\label{sec:forecasts}
\begin{figure}
\centering
\includegraphics[width=0.45\textwidth, trim = 0mm 0mm 0mm 0mm]{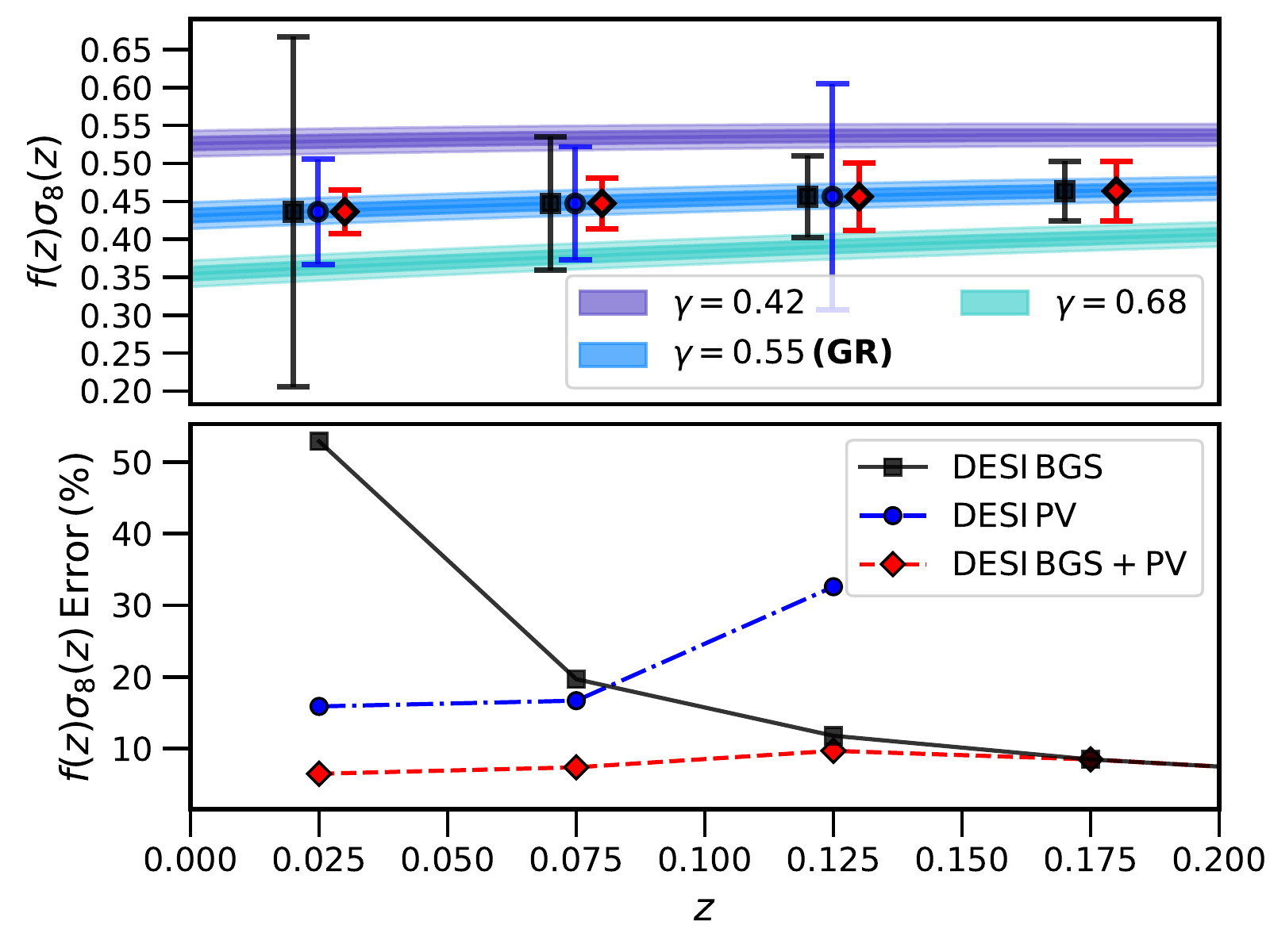}
\caption{Forecasts for constraints on the growth rate in $0.05$ redshift bins from the DESI BGS (black points), PV (blue points) and combined (red points) surveys. We assume clustering measurements are made up to $k_{\mathrm{max}}=0.2h\,\mathrm{Mpc^{-1}}$. The top panel shows the forecasts offset slightly from the mid-point of the bin for clarity, and centred on the GR prediction alongside different $\gamma$-parameterisations of gravity (coloured bands), where \citet{Planck2018} constraints on other cosmological parameters are used to produce the predictions for the growth rate. The bottom panel shows the relative error on the growth rate in each bin. The combined constraints are substantially improved with the inclusion of the PV sample compared to BGS alone, allowing for much greater distinction between different parameterisations of gravity at low redshift.}
\label{fig:forecasts}
\end{figure}

\begin{figure}
\centering
\includegraphics[width=0.45\textwidth, trim = 0mm 0mm 0mm 0mm]{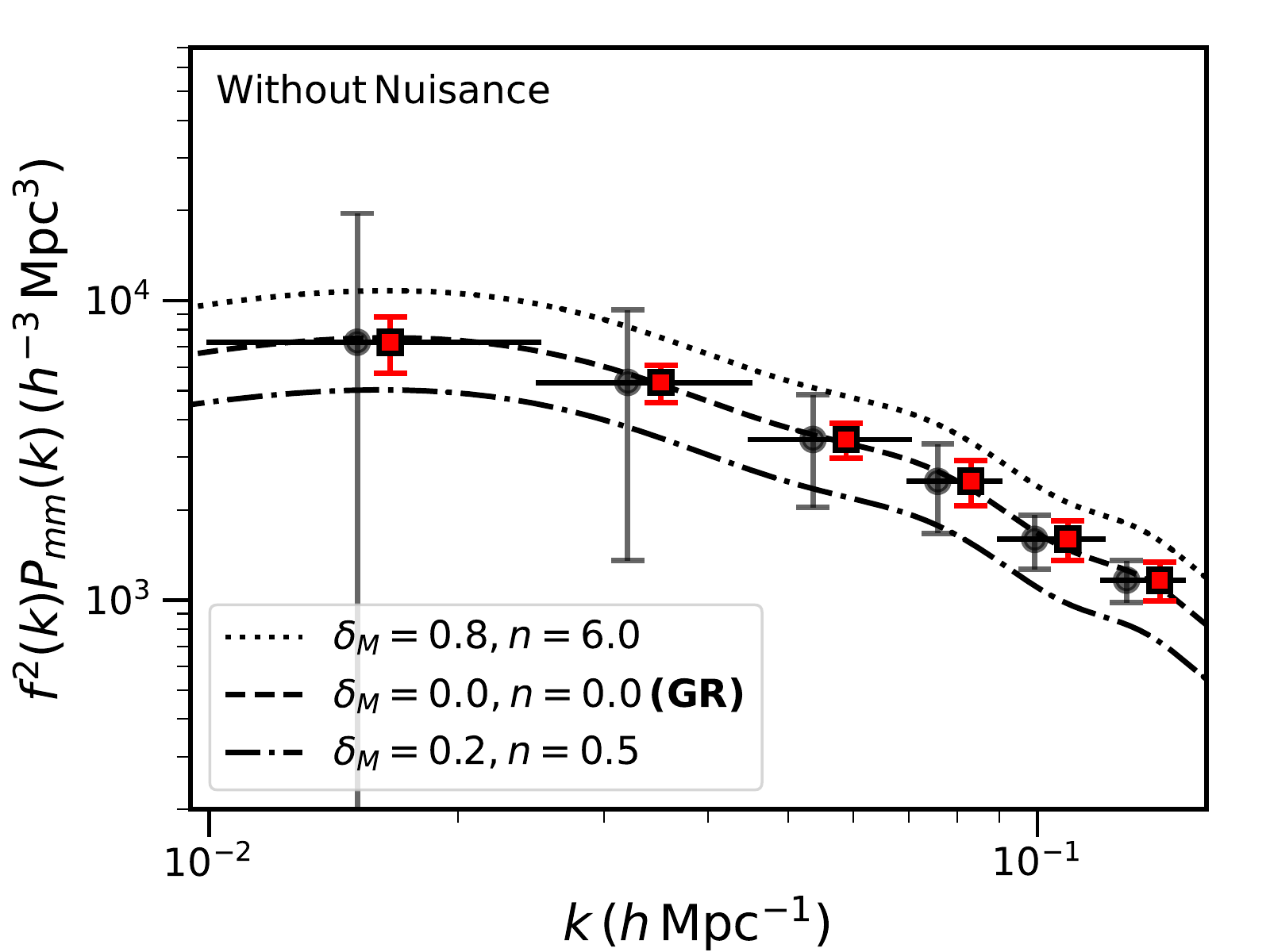}
\includegraphics[width=0.45\textwidth, trim = 0mm 0mm 0mm 0mm]{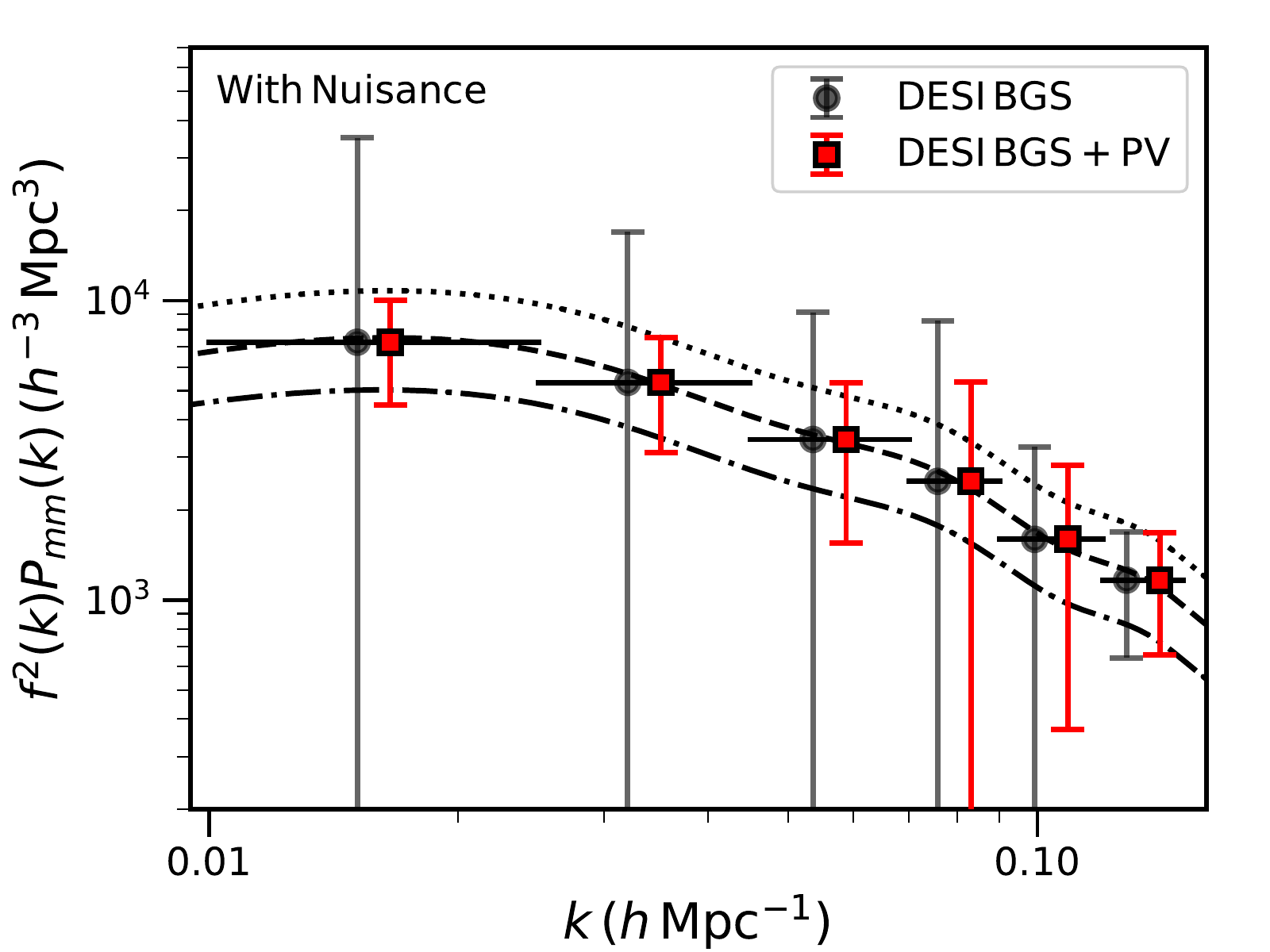}
\caption{Forecasts for measurements of the velocity power spectrum in different $k$-bins, allowing linear galaxy bias to vary independently in each bin. Points and errors show the forecasts, while the solid horizontal lines are the extent of each bin. The dashed/dotted lines are the predictions from theories of gravity with the same \citep{Planck2018} cosmological parameters but different effective Planck masses. Such measurements are expected to be sensitive to the scale-dependent damping of the galaxy density and velocity fields arising from non-linear redshift-space distortions, so we produce forecasts fixing these (top) and and allowing them to freely vary (bottom). In both cases, the inclusion of the PV sample greatly improves our ability to constrain the scale-dependence of the clustering.}
\label{fig:forecasts_sd}
\end{figure}

\begin{table}
    \centering
    \begin{tabular}{c|c|c|c}
    Redshift bin & DESI BGS & DESI PV & DESI BGS+PV \\ \hline \hline
       $0.00 \le z \le 0.05$ & 185.9 (52.9) & 26.9 (15.9) & 12.1 (6.5) \\ 
       $0.05 \le z \le 0.10$ & 69.1 (19.7) & 19.1 (16.7) & 10.4 (7.4) \\ 
       $0.10 \le z \le 0.15$ & 42.4 (11.8) & 34.1 (32.6) & 18.8 (9.7) \\ 
       $0.00 \le z \le 0.10$ & 65.9 (18.9) & 15.3 (10.8) & 6.5 (4.3) \\ 
       $0.00 \le z \le 0.15$ & 36.1 (10.1) & 13.8 (10.1) & 5.3 (3.7) \\ 
    \end{tabular}
    \caption{Forecasts for the percentage error on the growth rate of structure in various low redshift bins from DESI redshifts and PVs for $k_{\mathrm{max}}=0.1h\,\mathrm{Mpc^{-1}}$ ($k_{\mathrm{max}}=0.2h\,\mathrm{Mpc^{-1}}$). `DESI BGS' denotes forecasts using only the DESI-BGS redshifts to measure the galaxy-galaxy clustering, `DESI PV' are forecasts using only the PV sample (combining FP and TF samples) to measure the velocity-velocity clustering, and `DESI BGS+PV' shows constraints fitting both the auto and cross-correlations of galaxies and velocities simultaneously. In all cases we marginalise over all nuisance parameters identified in Section~\ref{sec:forecasts}.}
    \label{tab:forecasts}
\end{table}

The inclusion of PVs alongside redshifts in cosmological parameter estimates allows us to place tighter constraints on potential large-scale deviations from General Relativity (GR) and modified gravity theories. In this section, we take the anticipated sky area and redshift distribution of our FP and TF-based PV estimates and combine them with the expected redshift distribution for the DESI BGS sample. We then forecast the constraints on the growth of structure and a particular modified gravity model to demonstrate the power of this combined dataset.

Forecasts are produced using the Fisher matrix method of \cite{Howlett2017a,Howlett2017b}, which models the full suite of two-point correlations (power spectra) that can be measured from a partially overlapping set of redshifts and PVs. In paticular, we assume that one measures the auto- and cross-power spectrum from the available distribution of redshifts ($\delta$) and peculiar velocities ($v$). These power spectra models are given, as a function of separation $k$ and angle $\mu$, via
\begin{align}
    P_{\delta\delta}(k,\mu,z) &= (b\sigma_{8}+f\sigma_{8}\mu^{2})^{2}D^{2}_{g}(k,\mu)P_{mm}(k)/\sigma^{2}_{8} \notag \\
    P_{vv}(k,\mu,z) &= \biggl(\frac{H(z)}{1+z}f\sigma_{8}\mu^{2}\biggl)^{2}D^{2}_{u}(k)P_{mm}(k)/\sigma^{2}_{8} \notag \\
    P_{\delta v}(k,\mu,z) &= (b\sigma_{8}+f\sigma_{8}\mu^{2})\frac{H(z)}{1+z}\frac{f\sigma_{8}\mu}{k}D_{g}(k,\mu)D_{u}(k)P_{mm}(k)/\sigma^{2}_{8}.
    \label{eq:pvv}
\end{align}

These models include free parameters for linear galaxy bias, $b\sigma_{8}$, the normalised growth rate of structure $f\sigma_{8}$, and non-linear redshift space distortions in both the galaxy density and velocity fields. The latter are parameterised by the functions $D_{g}(k,\mu) = (1+1/2[k\mu\sigma_{g}]^{2})^{-1/2}$ and $D_{u}(k) = \mathrm{sinc}(k\sigma_{u})$ with free parameters $\sigma_{g}$ and $\sigma_{u}$ controlling the amount of damping in the non-linear power spectra. $P_{mm}$ is the underlying power spectrum of matter perturbations, which is the same in all three auto- and cross-spectra, highlighting that measuring and modelling all three can improve the constraints compared to measuring only redshifts or velocities.

We also account for noise in these measurements in the form of shot-noise (inversely proportional to the number density of galaxy redshifts or PV measurements), and PV uncertainties. The number densities for redshifts and PVs are taken from \cite{DESI_white_initial} and Figure~\ref{fig:DESInz} respectively and we assume the PV errors scale with comoving distance $r(z)$ as $\sigma^{2}_{\mathrm{obs}} = [400 r^{2}(z) + 300^{2}]\,\mathrm{km^{2}\,s^{-2}}$, which matches well the uncertainties seen in current FP and TF surveys. The code used to produce the cosmological forecasts in this paper is publicly available at \url{https://github.com/CullanHowlett/PV_fisher}.

Fisher matrix forecasts for combined redshift and PV surveys are weakly sensitive to the choice of fiducial/central parameters \citep{Koda2014,Howlett2017a}. In this work, we set values of $b(z) = 1.34D^{-1}(z)$, $f(z)\sigma_{8}(z) = \Omega^{0.55}_{m}(z)D(z)\sigma_{8,0}$, $\sigma_{g} = 4.24h^{-1}\,\mathrm{Mpc}$ and $\sigma_{u} = 13h^{-1}\,\mathrm{Mpc}$. These latter two values were found to reproduce the effects of non-linear RSD well in simulations \citep{Koda2014}. $D(z)$ is the linear growth factor in our fiducial cosmology. Our Fisher matrix forecasts are far more sensitive to the choice of measurement scales that we can assume to measure and model, so we present forecasts limited to mostly linear scales ($k_{\mathrm{max}}=0.1h\,\mathrm{Mpc}^{-1}$) and quasi-linear scales ($k_{\mathrm{max}}=0.2h\,\mathrm{Mpc}^{-1}$). The former of these values is somewhat pessimistic --- current joint analyses of the density and velocity fields can model information up to $k=0.2h\,\mathrm{Mpc}^{-1}$ with good systematic control (e.g., \citealt{Qin2019, Lai2022}), and we expect to be able to do the same with with DESI.

Assuming the redshift and velocity distributions are Gaussian, the three power spectra given in Eq.~\ref{eq:pvv} contain all available  cosmological information and so Fisher forecasts using these models provide an estimate of the tightest possible cosmological constraints. However, methods such as velocity and density field reconstruction \citep{Nusser2011,Boruah2019,Said:2020,Qin2023a} will also be investigated with the DESI PV data and could improve beyond these as they may access \textit{non-Gaussian} information. For the purposes of this work, we stick with the already compelling, simpler, Fisher forecasts but note that these could also end up being conservative.

Our forecasts for a $\sim$14 000 deg$^{2}$ DESI survey of redshifts and PVs for different low-redshift ranges are presented in Table~\ref{tab:forecasts} and plotted in Fig.~\ref{fig:forecasts}. The case for RSD-only can be compared to the BGS constraints in \cite{DESI_white_initial}, although the redshift-binning is different. When the same redshift bins are considered we find good agreement between the two sets of predictions, despite many differences in the forecasting model and methodology, which lends credence to our predictions. Overall, from the combination of redshifts and PVs and $k_{\mathrm{max}}=0.2h\,\mathrm{Mpc}^{-1}$, we predict two distinct $\sim$7\% measurements on the growth rate for $0.0 < z < 0.05$ and $0.05 < z < 0.1$, a $\sim$8$\times$ and $\sim$2.5$\times$ respective improvement over the BGS alone. Combining the information across the full $z < 0.15$ redshift range where both redshifts and PVs are expected to be measurable, we envision a $4\%$ measurement of the growth rate of structure, over two times better than achievable with redshifts only. Interestingly (assuming we use all data $0.0 < z < 0.15$ and fit up to a reasonable $k_{max}=0.2h\,\mathrm{Mpc}^{-1}$) we also find comparable, and equally tight, constraints on the growth rate of structure are possible treating the redshifts and PVs as independent tracers. This highlights potential opportunities to perform cross-validations of the two, or tests for new physics that would change only one or the other of these two measurements.

As shown in the top panel of Figure~\ref{fig:forecasts}, such tight growth rate constraints will provide improved distinction between different models or parameterisations of gravity, particularly in the redshift regime where we expect deviations between models to be greatest, or when combined with other higher redshift DESI measurements.  In addition, Fig.~\ref{fig:forecasts_sd} demonstrates that we may be able to constrain the scale-dependence of the growth rate of structure --- we show forecasts for the growth rate in specific $k$-ranges where for each $k$-bin we have allowed both the galaxy bias and growth rate to vary independently as a function of scale. In this case, non-linear RSD, which is a scale-dependent function, strongly affects our ability to obtain informative constraints. As such we produce forecasts with and without allowing $\sigma_{g}$ and $\sigma_{u}$ to also vary in each $k$-bin.

GR predicts a scale-\textit{independent} growth rate, and a departure from this as a function of $k$ can rule out GR and constrain alternative models. As an example we have also plotted (as dotted and dashed lines in Fig.~\ref{fig:forecasts_sd}) two Horndeski scalar-tensor gravitational models currently allowed by the combination of CMB+BAO data \citep{Noller2019}. Both models are characterised by deviations in the effective Planck mass via $M^{2}_{s} = M^{2}_{\mathrm{Pl}}(1+\delta_{M}a^{n})$, where $\delta_{M}$ and $n$ are free parameters, $a$ is the scale factor and $M_{\mathrm{Pl}}$ the standard Planck mass. On large scales, the strong degeneracy between galaxy bias and the growth rate combined with cosmic variance means that the BGS provides poor constraints on the scale dependence. However, the addition of the DESI PV survey makes this a viable and precise test and offers an excellent opportunity to test gravity in a novel way. The measurements including the DESI PV survey as shown in Fig.~\ref{fig:forecasts_sd} would potentially rule out both of these currently viable Horndeski models, but the significance of the results would depend strongly on our understanding of non-linear redshift-space distortions. Given that we have allowed the corresponding nuisance parameters to vary in each $k$-bin, rather than as a single parameter across the full power spectrum (like in our forecasts for different $z$-bins), it seems likely we can place some reasonable priors on the functional form of these such that the reality will lie somewhere between the forecasts presented in the two panels of Fig.~\ref{fig:forecasts_sd}. It may be possible to go even further using knowledge of galaxy evolution physics or assuming a redshift dependent model for the nuisance parameters and fitting alongside/extrapolating from the higher redshift DESI data.

\section{Summary and Conclusions}
\label{sec:conclusions}

This paper presents our program for the largest and most extensive PV survey to date. It uses both the FP and the TF relation as redshift-independent distance indicators. We have provided a detailed description of the selection criteria used to identify suitable targets for both distance indicators as well as our validation of these selection criteria. The complete target catalogues are provided on VizieR using the structure outline in Tables \ref{tab:target_fp} to \ref{tab:target_ext}. Additionally, we also provide supplementary catalogues for our targets that contain the parameters derived from the DESI Legacy Imaging Survey DR9 that were used for the selection of targets (see Tables \ref{tab:base_targets} to \ref{tab:flag_targets}) and their validation (see Tables \ref{tab:sga_targets} and \ref{tab:zoo_targets}) using SGA and GalaxyZoo. 

With 427 273 FP and 129 772 TF galaxy targets north of declination $-30$\textdegree~the DESI PV survey is the first large-scale survey to target both the FP and the TF-relation using the same instrument. Both target classes cover the same footprint and are identified using a consistent methodology, which will allow us to compare both distance indicators and better identify any biases. Based on our validation, we expect a success rate of our photometric classification of FP target of about $2/3^{\mathrm{rd}}$ and a success rate of our classification of TF targets of $~85\%$. Considering the fibre assignment of DESI as well as an estimate of the success rate of obtaining sufficient quality spectra based on early data from survey validation, we predict a final survey containing $\sim$133 000 FP galaxies and $\sim$53 000 TF galaxies within the DESI spectroscopic footprint limited to $z\lesssim0.15$.

The first results of the spectroscopic observations of our target galaxies that were carried out during the DESI survey validation will be provided in upcoming papers by Said et al (in preparation) for the FP and in Dougless et al (in preparation) for the TF relation. Once the survey is complete it will provide a notable improvement in the $f \sigma_{8}$ measurements obtained by DESI at low redshifts. Combined with the larger number of BGS redshifts in this regime we demonstrate we should achieve constraints on the time and scale evolution of the growth rate in the nearby universe that greatly surpass those obtainable with redshifts alone --- including a potential $4\%$ measurement on the growth rate of structure using the entire $z<0.15$ redshift range --- which will allow us to better distinguish between a range of models of that depart from GR (see Table \ref{tab:forecasts} and Figure \ref{fig:forecasts} for details). The combined DESI PV and redshifts surveys thereby provide a valuable contribution to cosmology in additional to furthering our understanding of matter distribution in the local universe. 


\section*{Acknowledgements}
C. Saulder acknowledges support from the National Research Foundation of Korea (NRF) through grant No. 2021R1A2C101302413 funded by the Korean Ministry of Education, Science and Technology (MoEST)

C. Howlett and K. Said are supported by the Australian Government through the Australian Research Council’s Laureate Fellowship funding scheme (project FL180100168).

This research is supported by the Director, Office of Science, Office of High Energy Physics of the U.S. Department of Energy under Contract No. DE–AC02–05CH11231, and by the National Energy Research Scientific Computing Center, a DOE Office of Science User Facility under the same contract; additional support for DESI is provided by the U.S. National Science Foundation, Division of Astronomical Sciences under Contract No. AST-0950945 to the NSF’s National Optical-Infrared Astronomy Research Laboratory; the Science and Technologies Facilities Council of the United Kingdom; the Gordon and Betty Moore Foundation; the Heising-Simons Foundation; the French Alternative Energies and Atomic Energy Commission (CEA); the National Council of Science and Technology of Mexico (CONACYT); the Ministry of Science and Innovation of Spain (MICINN), and by the DESI Member Institutions: \url{https://www.desi.lbl.gov/collaborating-institutions}.

The DESI Legacy Imaging Surveys consist of three individual and complementary projects: the Dark Energy Camera Legacy Survey (DECaLS), the Beijing-Arizona Sky Survey (BASS), and the Mayall z-band Legacy Survey (MzLS). DECaLS, BASS and MzLS together include data obtained, respectively, at the Blanco telescope, Cerro Tololo Inter-American Observatory, NSF’s NOIRLab; the Bok telescope, Steward Observatory, University of Arizona; and the Mayall telescope, Kitt Peak National Observatory, NOIRLab. NOIRLab is operated by the Association of Universities for Research in Astronomy (AURA) under a cooperative agreement with the National Science Foundation. Pipeline processing and analyses of the data were supported by NOIRLab and the Lawrence Berkeley National Laboratory. Legacy Surveys also uses data products from the Near-Earth Object Wide-field Infrared Survey Explorer (NEOWISE), a project of the Jet Propulsion Laboratory/California Institute of Technology, funded by the National Aeronautics and Space Administration. Legacy Surveys was supported by: the Director, Office of Science, Office of High Energy Physics of the U.S. Department of Energy; the National Energy Research Scientific Computing Center, a DOE Office of Science User Facility; the U.S. National Science Foundation, Division of Astronomical Sciences; the National Astronomical Observatories of China, the Chinese Academy of Sciences and the Chinese National Natural Science Foundation. LBNL is managed by the Regents of the University of California under contract to the U.S. Department of Energy. The complete acknowledgments can be found at \url{https://www.legacysurvey.org/.}

The Photometric Redshifts for the Legacy Surveys (PRLS) catalog used in this paper was produced thanks to funding from the U.S. Department of Energy Office of Science, Office of High Energy Physics via grant DE-SC0007914.

The authors are honored to be permitted to conduct scientific research on Iolkam Du’ag (Kitt Peak), a mountain with particular significance to the Tohono O’odham Nation.

Software: We extensively used the following python packages: \emph{astropy} \citep{astropy:2013,astropy:2018}, \emph{numpy} \citep{numpy}, and \emph{matplotlib} \citep{matplotlib}.

\section*{Data Availability}
All data points shown in the figures are available in a machine-readable form on \href{https://doi.org/10.5281/zenodo.7680931}{https://doi.org/10.5281/zenodo.7680931}. Additionally, all target tables from Appendix \ref{app:cats} are provided on Vizier: [add link once paper is accepted].


\bibliographystyle{mnras}
\bibliography{paper_DESI_PV_targets}

\appendix

\section{Preliminary BGS selection criteria}
\label{app:BGS}
We used preliminary BGS selection criteria as proposed by \citet{BGS_pre_target}, which was based on the DESI Legacy Imaging Survey DR8. While there were adjustments to these criteria \citep{DESI_target_BGS} prior to the launch of the main survey of DESI, most of them only affected the fainter end of the selection, which was below our limiting magnitude or the inclusion of SGA objects, we already considered independently. 

\begin{itemize}
	\item ((fibermag\_r  $<$ (5.1 + mag\_r)) AND (mag\_r $<=$ 17.8)) OR ((fibermag\_r  $<$ 22.9) AND (mag\_r $>$ 17.8 ) AND	(mag\_r $<$ 20 ))
 	\item ($-1$ $<$ mag\_g - mag\_r) AND (mag\_g - mag\_r $<$ 4) AND ($-1$ $<$ mag\_r - mag\_z) AND (mag\_r - mag\_z $<$ 4)
	\item NOBS\_g, NOBS\_r, NOBS\_z $>$ 0 
	\item FLUX\_g, FLUX\_r, FLUX\_z $>$ 0 
	\item (GAIA\_PHOT\_G\_MEAN\_MAG-mag\_r,noext $>$ 0.6) OR (GAIA\_PHOT\_G\_MEAN\_MAG = 0)
	\item MASKBITS 1 (bright stars (GAIA and Tycho)), 12 (SGA galaxies), and 13 (globular clusters) are zero 
	\item FRACMASKED\_g, FRACMASKED\_r, FRACMASKED\_z $<$ 0.4 
	\item FRACFLUX\_g, FRACFLUX\_r, FRACFLUX\_z $<$ 5
	\item FRACIN\_g, FRACIN\_r, FRACIN\_z $>$ 0.2 
\end{itemize}

The abbreviations used for the selection criteria are the following: The extinction-corrected magnitudes mag\_x and fiber magnitudes fibermag\_x are calculated using the various FLUX\_x and FIBERFLUX\_x respectively as well as the Milky Way transmission values (in the case of the usual values corrected for galactic extinction) MW\_TRANSMISSION\_X found in the DESI Legacy Imaging Survey catalogues using the following equation:
\begin{equation}
\textrm{mag}\_\textrm{x} = 22.5 - 2.5\ \textrm{log}_{10}\left(\frac{\textrm{FLUX}\_\textrm{X}}{\textrm{MW}\_\textrm{TRANSMISSION}\_\textrm{X}} \right).
\label{eq:mag}
\end{equation}
number of images that contribute to the central pixel: NOBS\_x; model flux: FLUX\_x; Gaia g band magnitude: GAIA\_PHOT\_G\_MEAN\_MAG; non-galactic extinction corrected r band model magnitude: mag\_r,noext; profile-weighted fraction of pixels masked from all observations of this object: FRACMASKED\_x; profile-weighted fraction of the flux from other sources divided by the total flux: FRACFLUX\_x; fraction of a source's flux within the blob\footnote{contiguous set of pixels associated with each source detection}: FRACIN\_g; fibre magnitude in the r band: fibermag\_r. For more details on the parameters, see \url{https://www.legacysurvey.org/dr9/catalogs/} and concerning the maskbits: \url{https://www.legacysurvey.org/dr9/bitmasks/}.

\section{Catalogues}
\label{app:cats}
This appendix contains the first five lines of all catalogue tables to outline their structure. The full catalogues are available on Vizier: [add Vizier link once paper is published].

The catalogue in Table \ref{tab:target_fp} contains the targets of our FP sample. As the fibres are only placed on the center of each target, it contains 427 273 fibre pointings in total. For the TF relation sample, we have 129 772 targets and 3 pointings for each of them, giving us the 389 316 entries in Table \ref{tab:target_tf}. Tables \ref{tab:target_sga} and \ref{tab:target_ext} also provide multiple pointings per target for our additional SGA and extend target samples respectively, that will be used for calibration. 

Our base sample obtained used for all specific target selections corresponds to a slightly modified version of the preliminary BGS selection of DESI \citep{BGS_pre_target}. For the 4 456 142 objects, we provide the basic parameters used for the section in Table \ref{tab:base_targets}. Derived photometric parameters are given in Table \ref{tab:photo_targets} and flags indicating samples and footprints can be found in Table \ref{tab:flag_targets}. 

We used the Siena Galaxy Atlas for calibrations as well as target selection. Therefore, we provide a list of all used parameters for the objects in the basic target selection and the SGA in Table \ref{tab:sga_targets}. Additionally, we also used GalaxyZoo to assess the quality of our selection and all necessary parameters to reproduce this selection are provided in Table \ref{tab:zoo_targets}.

The combination of OBJID, BRICKID, and BRICKNAME is a unique identifier that can be used to link all tables together and also back to the DESI Legacy Imaging Survey DR9 \citep{DESI_Imaging}. 
\begin{table*}
\begin{center}
\begin{tabular}{ccc|ccccccc}
OBJID & BRICKID & BRICKNAME & RA & DEC & TYPE & SERSIC & Z\_PHOT\_MEDIAN & Z\_PHOT\_L95 & SHAPE\_R \\
\hline
62 & 517233 & 2265p342 & 226.4355 & 34.3231 & REX & 1.0 & 0.1391 & 0.0613 & 1.0949 \\
99 & 517233 & 2265p342 & 226.4384 & 34.2614 & SER & 2.75 & 0.1043 & 0.0584 & 0.5902 \\
266 & 517233 & 2265p342 & 226.4516 & 34.1455 & SER & 0.9 & 0.1206 & 0.0729 & 1.7629 \\
340 & 517233 & 2265p342 & 226.4561 & 34.3241 & EXP & 1.0 & 0.1008 & 0.0471 & 2.1094 \\
508 & 517233 & 2265p342 & 226.4674 & 34.1878 & SER & 2.94 & 0.1238 & 0.085 & 4.5398 \\
\end{tabular}
\caption{Basic parameters of our basic sample. OBJID, BRICKID, and BRICKNAME: identifiers from the DESI Legacy Imaging Survey DR9; RA and DEC: equatorial coordinates in degree, TYPE: model used for best fit (PSF: point spread function, REX: round exponential galaxy model, DEV: de Vaucouleurs model, EXP: exponential model, SER: Sersic model); SERSIC: S\'ersic index of the best fit model; Z\_PHOT\_MEDIAN: median value of the photometric redshift estimate; Z\_PHOT\_L95: lower $2\,\sigma$ limit of photometric redshift estimate; SHAPE\_R: uncorrected radius derived using the best fit model, in arcseconds.} \label{tab:base_targets} \end{center}
\end{table*}
\begin{table*}
\begin{center}
\begin{tabular}{ccc|ccccccccc}
OBJID & BRICKID & BRICKNAME & mag\_g & mag\_r & mag\_z & mag\_g\_err & mag\_r\_err & mag\_z\_err & BA\_ratio & circ\_radius & pos\_angle \\
\hline
62 & 517233 & 2265p342 & 20.398 & 19.957 & 19.631 & 0.012 & 0.014 & 0.014 & 1.0 & 1.0949 & 0.0 \\
99 & 517233 & 2265p342 & 19.409 & 18.738 & 18.265 & 0.004 & 0.004 & 0.003 & 0.762 & 0.5151 & -25.9 \\
266 & 517233 & 2265p342 & 19.203 & 18.499 & 17.83 & 0.005 & 0.005 & 0.003 & 0.267 & 0.911 & -49.7 \\
340 & 517233 & 2265p342 & 19.964 & 19.661 & 19.421 & 0.01 & 0.013 & 0.014 & 0.46 & 1.4304 & 68.8 \\
508 & 517233 & 2265p342 & 19.137 & 18.134 & 17.304 & 0.007 & 0.005 & 0.003 & 0.263 & 2.3282 & 79.9 \\
\end{tabular}
\caption{Derived photometric parameters of our basic sample. OBJID, BRICKID, and BRICKNAME: identifiers from the DESI Legacy Imaging Survey DR9; mag\_g, mag\_r, and mag\_z: apparent magnitudes in the g, r, and z band respectively (corrected for Milky Way dust extinction);  mag\_g\_err, mag\_r\_err, and mag\_z\_err: error associated with the apparent magnitudes; BA\_ratio: ratio between the semi-minor and semi-major axis derived using the best fit model; circ\_radius: circularised radius in arcseconds; pos\_angle: position angle (in degree) derived using the best fit model.} \label{tab:photo_targets} \end{center}
\end{table*}
\begin{table*}
\begin{center}
\begin{tabular}{ccc|ccccccccc}
OBJID & BRICKID & BRICKNAME & etg\_flag & ltg\_flag & inSGA & inBGS\_old & inbright & is\_lowz & in\_spec & in\_sdssdr7 & in\_galaxyzoo \\
\hline
62 & 517233 & 2265p342 & 0 & 0 & 0 & 1 & 0 & 1 & 1 & 1 & 0 \\
99 & 517233 & 2265p342 & 0 & 0 & 0 & 1 & 0 & 1 & 1 & 1 & 0 \\
266 & 517233 & 2265p342 & 0 & 0 & 0 & 1 & 0 & 1 & 1 & 1 & 0 \\
340 & 517233 & 2265p342 & 0 & 0 & 0 & 1 & 0 & 1 & 1 & 1 & 0 \\
508 & 517233 & 2265p342 & 0 & 0 & 0 & 1 & 0 & 1 & 1 & 1 & 0 \\
\end{tabular}
\caption{Flags of our basic sample. OBJID, BRICKID, and BRICKNAME: identifiers from the DESI Legacy Imaging Survey DR9; etg\_flag: flag indicating if in FP sample; ltg\_flag: flag indicating if in TF sample; inSGA: flag indicating if can be found in SGA; inBGS\_old: flag indicating if can be found in the preliminary BGS target selection of DESI; inbright: flag indicating if brighter than 18 mag in r band; is\_lowz: flag indicating if photometric redshifts are within the range set by the FP sample selection; in\_spec: flag indicating if target is within range of planned DESI spectroscopic tiles; in\_sdssdr7: flag indicating if target is within the North Galactic Cap of the SDSS DR7 footprint; in\_galaxyzoo: flag indicating if target is in GalaxyZoo} \label{tab:flag_targets} \end{center}
\end{table*}
 
\begin{table*}
\begin{center}
{\scriptsize
\begin{tabular}{ccc|cccccccc}
OBJID & BRICKID & BRICKNAME & RA & DEC & REF\_EPOCH & OVERRIDE & PVTYPE & PVPRIORITY & POINTINGID & TARGETID \\
\hline
1890 & 517233 & 2265p342 & 226.5663 & 34.2078 & 2015.5 & 0 & FPT & 1 & 1 & 2843307478614018 \\
1989 & 517233 & 2265p342 & 226.5727 & 34.2791 & 2015.5 & 0 & FPT & 1 & 1 & 2843307478614016 \\
3778 & 517233 & 2265p342 & 226.7099 & 34.364 & 2015.5 & 0 & FPT & 1 & 1 & 2843307478614017 \\
210 & 520796 & 2267p350 & 226.6193 & 34.8929 & 2015.5 & 0 & FPT & 1 & 1 & 2843322422919168 \\
969 & 514840 & 2263p337 & 226.2728 & 33.6786 & 2015.5 & 0 & FPT & 1 & 1 & 2843297441644544 \\
\end{tabular}
}
\caption{Fibre pointings for the FP targets. OBJID, BRICKID, and BRICKNAME: identifiers from the DESI Legacy Imaging Survey DR9; RA and DEC: equatorial coordinates in degree, REF\_EPOCH: reference epoch for the coordinates; OVERRIDE: flag for the DESI targeting pipeline; PVTYPE: type of the PV target; PVPRIORITY: relative priority of the specific pointing; POINTINGID: Pointing ID to distinguish different pointings for the same target galaxy.} \label{tab:target_fp} \end{center}
\end{table*}
\begin{table*}
\begin{center}
{\scriptsize
\begin{tabular}{ccc|cccccccc}
OBJID & BRICKID & BRICKNAME & RA & DEC & REF\_EPOCH & OVERRIDE & PVTYPE & PVPRIORITY & POINTINGID & TARGETID \\
\hline
262 & 520796 & 2267p350 & 226.622 & 34.9823 & 2015.5 & 1 & TFT & 1 & 1 & 2843322422919169 \\
262 & 520796 & 2267p350 & 226.6232 & 34.981 & 2015.5 & 0 & TFT & 2 & 2 & 2847720469430272 \\
262 & 520796 & 2267p350 & 226.6244 & 34.9797 & 2015.5 & 1 & TFT & 1 & 3 & 2843322422919170 \\
2738 & 514840 & 2263p337 & 226.4014 & 33.6854 & 2015.5 & 1 & TFT & 1 & 1 & 2843297441644545 \\
2738 & 514840 & 2263p337 & 226.4043 & 33.6845 & 2015.5 & 0 & TFT & 2 & 2 & 2847695488155648 \\
\end{tabular}
}
\caption{Fibre pointings for the TF relation targets. Columns: same as Table \ref{tab:target_fp}.} \label{tab:target_tf} \end{center}
\end{table*}
\begin{table*}
\begin{center}
{\scriptsize
\begin{tabular}{ccc|cccccccc}
OBJID & BRICKID & BRICKNAME & RA & DEC & REF\_EPOCH & OVERRIDE & PVTYPE & PVPRIORITY & POINTINGID & TARGETID \\
\hline
3683 & 517233 & 2265p342 & 226.7036 & 34.2186 & 2015.5 & 0 & SGA & 3 & 1 & 2852103571636224 \\
3703 & 514840 & 2263p337 & 226.4735 & 33.7726 & 2015.5 & 0 & SGA & 3 & 1 & 2852093534666752 \\
307 & 518425 & 2267p345 & 226.6146 & 34.4591 & 2015.5 & 0 & SGA & 3 & 1 & 2852108571246593 \\
2608 & 518425 & 2267p345 & 226.7694 & 34.6055 & 2015.5 & 0 & SGA & 3 & 1 & 2852108571246592 \\
3677 & 516039 & 2268p340 & 226.8813 & 34.074 & 2015.5 & 0 & SGA & 3 & 1 & 2852098563637248 \\
\end{tabular}
}
\caption{Fibre pointings for the Siene Galaxy Atlas targets. Columns: same as Table \ref{tab:target_fp}.} \label{tab:target_sga} \end{center}
\end{table*}
\begin{table*}
\begin{center}
{\scriptsize
\begin{tabular}{ccc|cccccccc}
OBJID & BRICKID & BRICKNAME & RA & DEC & REF\_EPOCH & OVERRIDE & PVTYPE & PVPRIORITY & POINTINGID & TARGETID \\
\hline
3253 & 540249 & 1936p392 & 193.7728 & 39.2474 & 2015.5 & 1 & EXT & 3 & 1 & 2852200107737094 \\
3253 & 540249 & 1936p392 & 193.7733 & 39.2423 & 2015.5 & 1 & EXT & 3 & 2 & 2852200107737090 \\
3253 & 540249 & 1936p392 & 193.7737 & 39.2372 & 2015.5 & 1 & EXT & 3 & 3 & 2852200107737097 \\
3253 & 540249 & 1936p392 & 193.7742 & 39.2321 & 2015.5 & 1 & EXT & 3 & 4 & 2852200107737088 \\
3253 & 540249 & 1936p392 & 193.7746 & 39.227 & 2015.5 & 1 & EXT & 3 & 5 & 2852200107737091 \\
\end{tabular}
}
\caption{Fibre pointings for the extended targets. Columns: same as Table \ref{tab:target_fp}.} \label{tab:target_ext} \end{center}
\end{table*}

\begin{table*}
\begin{center}
{\scriptsize
\begin{tabular}{ccc|ccccccccc}
OBJID & BRICKID & BRICKNAME & SGA\_ID & PA & BA & G\_MAG\_SB25 & R\_MAG\_SB25 & Z\_MAG\_SB25 & RADIUS\_SB25 & DIAM & MORPHTYPE \\
\hline
1890 & 517233 & 2265p342 & 840129 & 6.1 & 0.354 & -1.0 & -1.0 & -1.0 & -1.0 & 0.4863 & E? \\
1989 & 517233 & 2265p342 & 116703 & 65.5 & 0.8671 & 17.549 & 16.515 & 15.976 & 11.1289 & 0.5128 & E \\
3683 & 517233 & 2265p342 & 917325 & 171.9 & 0.9304 & 16.829 & 16.303 & 15.992 & 11.244 & 0.493 & Sbc \\
3778 & 517233 & 2265p342 & 1378051 & 57.8 & 0.8225 & 17.27 & 16.252 & 15.629 & 11.6505 & 0.5438 & E \\
262 & 520796 & 2267p350 & 1006008 & 142.6 & 0.2874 & 18.105 & 17.107 & 16.428 & 11.3469 & 0.4849 & E? \\
\end{tabular}
}
\caption{Parameters used from the Siena Galaxy Atlas matched to our basic sample. OBJID, BRICKID, and BRICKNAME: identifiers from the DESI Legacy Imaging Survey DR9; SGA\_ID: identifier within the Siena Galaxy Atlas; PA: position angle (in degree) in SGA; BA: axis-ratio in SGA; G\_MAG\_SB25, R\_MAG\_SB25, and Z\_MAG\_SB25: cumulative brightness measured within the 25\,mag isophote for the g, r, and z band, in magnitudes, respectively; RADIUS\_SB25: radius corresponding to the 25\,mag isophote; DIAM: diameter in arcminutes of the 26\,mag isophote; MORPHTYPE: morphological type according to visual inspection.} \label{tab:sga_targets} \end{center}
\end{table*}

\begin{table*}
\begin{center}
\begin{tabular}{ccc|cccc}
\hline
OBJID & BRICKID & BRICKNAME & p\_el\_debiased & p\_cs\_debiased & nvote \\
1297 & 517233 & 2265p342 & 0.207 & 0.735 & 21 \\
1890 & 517233 & 2265p342 & 0.411 & 0.515 & 28 \\
1989 & 517233 & 2265p342 & 0.701 & 0.266 & 65 \\
2541 & 517233 & 2265p342 & 0.083 & 0.298 & 35 \\
3683 & 517233 & 2265p342 & 0.0 & 1.0 & 28 \\
\end{tabular}
\caption{Parameters used from the GalaxyZoo matched to our basic sample. BJID, BRICKID, and BRICKNAME: identifiers from the DESI Legacy Imaging Survey DR9; p\_el\_debiased: debiased fraction of votes on elliptical morphology; p\_cs\_debiased: debiased fraction of votes on spiral morphology; nvote: total number of votes for that object.} \label{tab:zoo_targets} \end{center}
\end{table*}


\bsp	
\label{lastpage}
\end{document}